%% file: main.tex
\documentclass{article}

\usepackage{PRIMEarxiv}

\usepackage[utf8]{inputenc} 
\usepackage[T1]{fontenc}    
\usepackage{hyperref}       
\usepackage{url}            
\usepackage{booktabs}       
\usepackage{amsfonts}       
\usepackage{nicefrac}       
\usepackage{microtype}      
\usepackage{lipsum}
\usepackage{fancyhdr}       
\usepackage{graphicx}       
\usepackage{amsmath, amsfonts} 
\usepackage{amssymb}
\usepackage{siunitx}
\usepackage{orcidlink}
\usepackage{acronym}
\graphicspath{{media/}}     

\title{A Guide to Bayesian Optimization in Bioprocess Engineering}

\author{
  Maximilian Siska \\
  IBG-1: Biotechnology \\
  Forschungszentrum Jülich \\
  Computational Systems Biotechnology \\
  RWTH Aachen University \\
  \And
  Emma Pajak \\
  The Sargent Centre for \\
  Process Systems Engineering \\
  Department of Chemical Engineering \\
  Imperial College London \\
  \And
  Katrin Rosenthal \\
  School of Science \\
  Constructor University Bremen \\
  \And
  Antonio del Rio Chanona \\
  The Sargent Centre for \\
  Process Systems Engineering \\
  Department of Chemical Engineering \\
  Imperial College London \\
  \And
  Eric von Lieres \\
  IBG-1: Biotechnology \\
  Forschungszentrum Jülich \\
  Computational Systems Biotechnology \\
  RWTH Aachen University \\
  \And
  Laura M. Helleckes \\
  Department of Chemical Engineering\\
  I-X Centre for AI in Science\\
  Imperial College London \\
  \texttt{l.helleckes@imperial.ac.uk} \\
}

\begin{document}
\maketitle

\begin{abstract}
Bayesian optimization has become widely popular across various experimental sciences due to its favorable attributes: it can handle noisy data, perform well with relatively small datasets, and provide adaptive suggestions for sequential experimentation.
While still in its infancy, Bayesian optimization has recently gained traction in bioprocess engineering.
However, experimentation with biological systems is highly complex and the resulting experimental uncertainty requires specific extensions to classical Bayesian optimization.
Moreover, current literature often targets readers with a strong statistical background, limiting its accessibility for practitioners.\\
In light of these developments, this review has two aims: first, to provide an intuitive and practical introduction to Bayesian optimization; and second, to outline promising application areas and open algorithmic challenges, thereby highlighting opportunities for future research in machine learning.
\end{abstract}


\input{01_Introduction/intro}
\input{02_Methodological_Background/1_bo_context}
\input{02_Methodological_Background/2_experiments}
\input{02_Methodological_Background/3_surrogate_model}
\input{02_Methodological_Background/4_opt_acquisition_fn}
\input{02_Methodological_Background/5_termination_criterion}
\input{03_Practical_Guidelines/1_case_study_context}
\input{03_Practical_Guidelines/2_experiments}
\input{03_Practical_Guidelines/3_surrogate_model}
\input{03_Practical_Guidelines/4_acquisition_function}
\input{04_Applications_Bioengineering/applications}
\input{05_Advanced_BO/advanced_bo}
\input{06_Reproducibility_and_Data_Deposition/data_deposition}
\input{07_Conclusions/conclusions}
\input{08_Boxes/background_gps}
\input{08_Boxes/hybrid_model_pymc}
\clearpage
\input{09_Glossary/glossary}
\section*{Acronyms}
\begin{acronym}
\acro{bo}[BO]{Bayesian optimization}
\acro{doe}[DoE]{Design of Experiments}
\acro{ei}[EI]{expected improvement}
\acro{gp}[GP]{Gaussian process}
\acro{mfbo}[MF-BO]{multi-fidelity Bayesian optimization}
\acro{ucb}[UCB]{upper confidence bound}
\end{acronym}
\acrodefplural{gp}[GPs]{Gaussian processes}
\acrodefplural{mogp}[GPs]{multi-output Gaussian processes}

\section*{Data and code availability}
Accompanying code, notebooks and figures are published in an accompanying GitHub repository~\cite{siskaLhelleckesBO_Empirical_ExamplesV0102025}.

\bibliographystyle{unsrt}  
\bibliography{references}

\end{document}

%% file: 01_Introduction/intro.tex
\section{Bayesian Optimization--From Decision Theory to Practical Experimental Design in Bioprocess Engineering}
\label{sec:intro}
Across various experimental sciences, \ac{bo} is increasingly recognized as a valuable tool for tackling complex, costly, and time-consuming optimization problems--conditions that are also common in bioprocess engineering.
While \ac{bo} saw a major increase in applications over the past decade, particularly in the context of \textbf{Machine Learning}~\cite{shahriariTakingHumanOut2016} (see Glossary in Section~\ref{sec:glossary}), its roots lie deep in the history of decision theory and experimental design, beginning in the 20th century.
\\
Major components of the modern \ac{bo} framework were already introduced in the context of sequential experimental design as early as the 1960s.
In two seminal papers~\cite{kushnerVersatileStochasticModel1962, kushnerNewMethodLocating1964}, applied mathematician Harold Kushner introduced the optimization of a noisy \textbf{objective function} using probabilistic models, specifically what we now call \textbf{\acp{gp}}, to represent the \textbf{uncertainty} about how different inputs affect an experimental outcome.
He also proposed heuristic strategies for selecting the next experiment, based on either the probability of improvement or how uncertain the outcome in a promising region is.
These concepts laid the foundation for the modern \ac{bo} approaches we apply today.
Notably, the field of geostatistics also promoted the use of \acp{gp} for experimental design as early as the 1950s, then promoted under the name of kriging after early work by Danie Krige~\cite{krigeStatisticalApproachMine1951} and Georges Matheron~\cite{matheronPrinciplesGeostatistics1963}.
\\
In the following decades, several scientists such as Vydūnas Šaltenis~\cite{saltenisMethodMultiextremalOptimization1971} and Jonas Močkus~\cite{mockusBayesianMethodsSeeking1975, mockusBayesianApproachGlobal1989} formalized and further promoted the concept of \ac{bo}, for example introducing \textit{expected improvement} (Section~\ref{sec:acquisition}) as a concept to design the next experiment, which is still frequently applied today.
By the 1990s, \ac{bo} had made the full transition from theory to practice~\cite{jonesEfficientGlobalOptimization1998}.
However, limited computational power still hindered its widespread adoption.
This changed drastically following the landmark study by Snoek et al.~\cite{snoekPracticalBayesianOptimization2012} in 2012, who demonstrated that \ac{bo} could outperform manual tuning and other automated methods for \textbf{hyperparameter} optimization in machine learning.
This success catalyzed widespread adoption in machine learning and sparked broader interest across scientific disciplines.
During this period, \ac{bo} was also rediscovered in the context of sequential experimental design~\cite{shahriariTakingHumanOut2016, frazierBayesianOptimizationMaterials2016,marchantBayesianOptimisationIntelligent2012, negoescuKnowledgeGradientAlgorithmSequencing2011}.
Notably, materials science and chemical engineering have emerged as important application domains for \ac{bo} in experimental optimization.
Researchers in these domains have leveraged \ac{bo} to guide discovery of materials~\cite{deshwalBayesianOptimizationNanoporous2021, diwaleBayesianOptimizationMaterial2022, jinBayesianOptimisationEfficient2023}, tune reaction conditions~\cite{shieldsBayesianReactionOptimization2021, taylorAcceleratedChemicalReaction2023}, and optimize reactor designs~\cite{parkMultiobjectiveBayesianOptimization2018, savageMultifidelityDatadrivenDesign2023}--challenges that mirror those encountered in biotechnology.
\\
In recent years, \ac{bo} has gained increasing attention in biotechnology, with applications ranging from drug discovery to protein design and reaction engineering.
Large datasets, such as those in clinical drug development, benefit from Bayesian approaches that not only yield statistical insights but also support planning, analysis, and decision-making~\cite{rubergApplicationBayesianApproaches2023}.
Furthermore, \ac{bo} has efficiently guided iterative robotic experiments in protein engineering~\cite{huProteinEngineeringBayesian2023}, the selection of production strains~\cite{helleckesHighthroughputScreeningCatalytically2024}, or the optimization of production media~\cite{freierFrameworkKrigingbasedIterative2016} and reaction conditions~\cite{pandiVersatileActiveLearning2022}. 
However, \ac{bo} is by no means limited to large datasets and automated experiments; it also excels in small-data settings.
In bioprocess engineering, the number of experiments that can be carried out might be limited due to costs, complexity, or the fact that automated laboratory equipment may be not available.
In such cases, \ac{bo} can perform effectively with well under 100~experiments, as demonstrated in the optimization of growth and production media~\cite{yoshidaHighThroughputOptimization2023, narayananAcceleratingCellCulture2025}, reaction engineering~\cite{ siedentopAvoidingReplicatesBiocatalysis2025,rosaMaximizingMRNAVaccine2022}, process engineering~\cite{hernandezrodriguezDesigningRobustBiotechnological2022} and downstream processing~\cite{japelBayesianOptimizationUsing2022,freierMultiobjectiveGlobalOptimization2017,narayananDesignBiopharmaceuticalFormulations2021}.
\\
Thus, \ac{bo} is becoming a promising tool for bioprocess engineering, offering data-efficient and robust optimization.
However, despite several reported case studies (summarized, e.\,g., in Gisperg et al.~\cite{gispergBayesianOptimizationBioprocess2025}), \ac{bo} remains underexplored in this field. 
On the one hand, the field presents unique challenges, such as complex measurement systems with varying biases  and high uncertainty resulting from the use of living organisms or parts thereof.
The models and algorithms needed to address these challenges are often not yet included in standard \ac{bo} software or documentation, creating an additional entry barrier for bioprocess engineers.
On the other hand, much of the existing literature is targeted at machine learning researchers and scientists from other domains.
\\
This review aims to comprehensively address these gaps, providing guidance on how to tackle challenging experimental systems in bioprocess engineering.
Our goals are twofold.
First, we offer the bioprocess engineering community an accessible methodological foundation and practical guidelines for addressing complex optimization problems.
Second, we highlight the specific characteristics of bioprocess engineering applications that present challenging and underexplored problems to the machine learning community.

%% file: 02_Methodological_Background/1_bo_context.tex
\section{Methodological Background of the Bayesian Optimization Framework}
\label{sec:method_background}
\begin{figure}[ht!]
    \centering
    \includegraphics[width=0.6\textwidth]{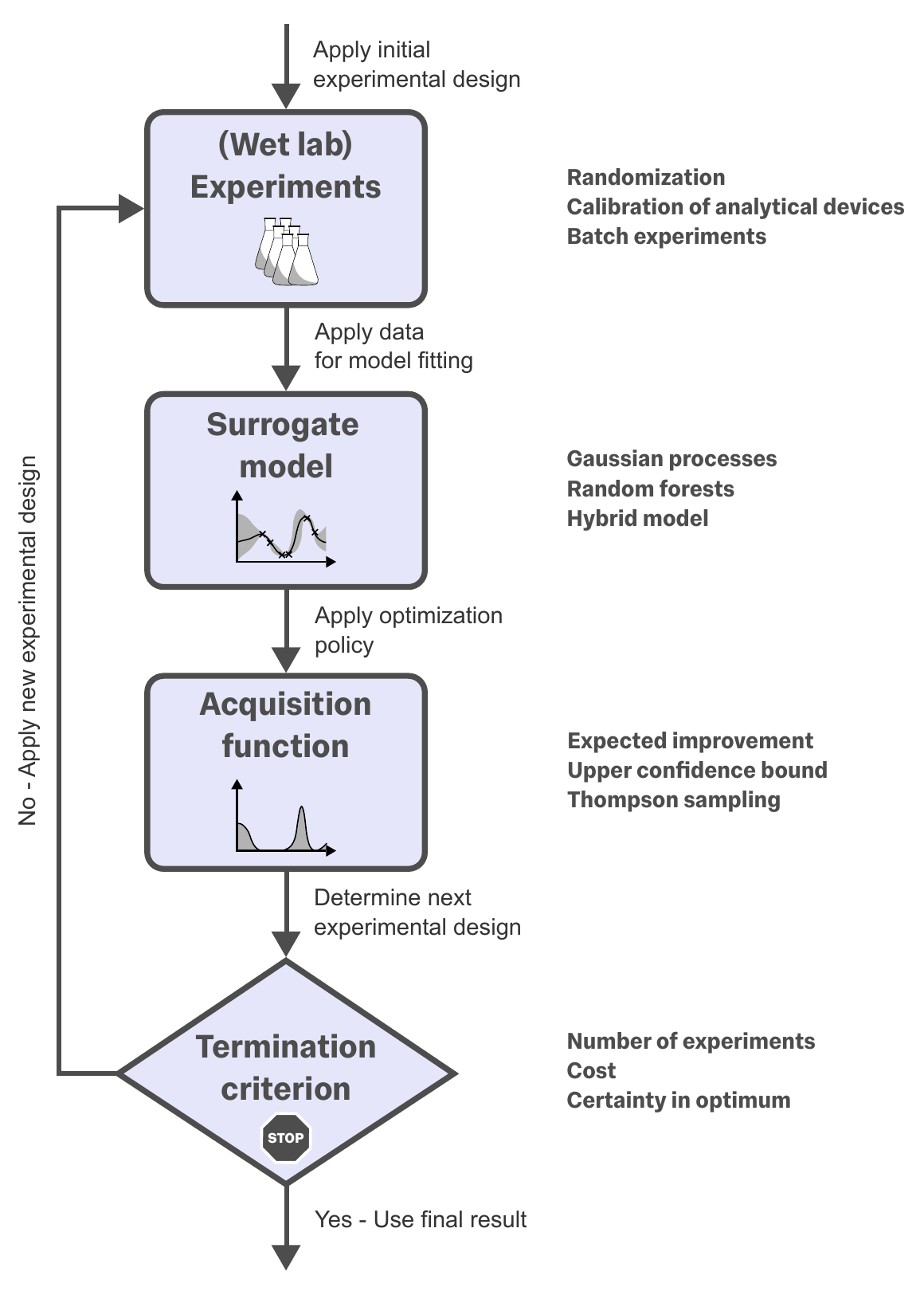}
    \caption{Overview of the four steps in the \ac{bo} workflow.\\
    \ac{bo} is a sequential, iterative process. Before the iterations begin, initial experiments are performed to obtain foundational observations. This step is typically conducted using space-filling designs such as Sobol sequences or Latin hypercube sampling. For the experiments themselves, important considerations include randomization, calibration of analytical devices to avoid biases, and distributing experiments across multiple batches (Section~\ref{sec:experiments}). Next, the generated data is used to fit a surrogate model, most commonly a \acs{gp} (Section~\ref{sec:gps}), although alternative methods can also be employed, such as decision trees or hybrid models that combine mechanistic and data-driven approaches (Box~\ref{box:hybrid_model}). Based on the surrogate model, the most informative next experimental conditions are selected using an acquisition function, depending on the chosen optimization policy. Common choices include expected improvement, upper confidence bound, or Thompson sampling (Section~\ref{sec:acquisition}). This procedure is then iteratively repeated by applying the selected experimental designs in the next round of experimentation. The iterative procedure is terminated based on a stopping criterion, such as the certainty with which the optimum has been identified, a predefined maximum number of experiments, or overall resource constraints.}
    \label{fig:bo-overview}
\end{figure}

In the field of bioprocess engineering and its related experimental work, we face many challenges that can be tackled as optimization problems, for example, finding the composition of a growth medium that yields the highest growth, or choosing between different biocatalysts for a bioprocess to obtain the highest reaction rate.
From a mathematical point of view, this is generalized in the field of optimization, where we seek to find the optimum of an objective function (i.\,e., our optimization target), which we evaluate at certain points by making one or more \textbf{observations} (i.\,e.,~conducting \textbf{experiments}).
In an iterative process, we decide where to perform these experiments by applying an \textbf{optimization policy} to the experimental data observed so far.
This policy can take many forms, ranging from naively picking random points within the \textbf{parameter space} to strategies that identify the next experimental conditions with the greatest potential to improve upon current results.
\\
\ac{bo} is a probabilistic, iterative method for identifying the optimum of a black-box objective function, meaning that we do not know the functional form of the optimization target a priori.
In bioprocess engineering, the objective function could be, for example, the titer of a product or the overall cost of a bioprocess.
As shown in Figure~\ref{fig:bo-overview}, \ac{bo} requires four elements:
\begin{enumerate}
\item wet-lab (or in silico) experiments (i.\,e., the data-generating system for the observations),
\item a \textbf{surrogate model} describing the black-box objective function,
\item an \textbf{acquisition function} (i.\,e., the mathematical function that assesses the value of potential new experiments according to the optimization policy), and
\item a termination criterion.
\end{enumerate}

\noindent In the following, each of these components will be explained in a dedicated subsection.
\clearpage

%% file: 02_Methodological_Background/2_experiments.tex
\subsection{Experiments}
\label{sec:experiments}
The first step in \ac{bo} is to conduct initial experiments, yielding a set of observations that can be used to fit the first surrogate model.
While the acquisition function (Section~\ref{sec:acquisition}) determines the design of subsequent experiments at later stages of the loop, the initial experiments, sometimes also referred to as excitation design, depend on the specific case study.
For example, prior knowledge of the optimization problem might influence the choice of experiments, i.\,e.~how to vary the \textbf{design variables} initially.
In general, space-filling designs have proven beneficial in many cases~\cite{jonesEfficientGlobalOptimization1998}.
These experimental designs aim to strategically cover the entire parameter space spanned by the design variables, using Sobol sequences or Latin hypercube sampling.
Other typical excitation designs follow established strategies from \ac{doe}, such as factorial designs.\\
As with classical \ac{doe}, experimental considerations such as randomization of samples, distribution of experiments across \textbf{batches} and minimization of nuisance factors play an important role in obtaining high-quality datasets.
This is especially important in bioprocess engineering, where positional biases (e.\,g., in microtiter plates) and batch-to-batch variability can significantly affect results.
These effects are therefore discussed in more detail in Section~\ref{sec:challenges_experiment}.
Beyond the experimental designs themselves, characteristics of the experimental system, for example the measurement noise of different analytical devices, can also be taken into account in \ac{bo}.
Many traditional \ac{doe} methods assume homoscedastic noise, i.\,e.,~constant variance across the full range of design variables.
However, \ac{bo} can also accommodate more complex scenarios with \textbf{heteroscedastic noise} in the observations, i.\,e.,~variance that changes with the design variables.
To identify the relationship between the measured quantities and the output of the analytical device, calibration can be performed using known reference standards to characterize the response of the devices.
Further details on how this information can be used in \ac{bo} will be discussed in Section~\ref{sec:challenges_experiment}.

%% file: 02_Methodological_Background/3_surrogate_model.tex
\subsection{Surrogate Model}
With the initial set of experiments complete, the next step in the \ac{bo} workflow (Figure~\ref{fig:bo-overview}) is to fit the surrogate model.
A surrogate model is a mathematical approximation that captures how the objective function (e.\,g., reaction yield or selectivity) changes in response to different design variables (e.\,g., temperature, pH).
In other words, it provides a mapping from the experimental inputs to the system’s output.
Surrogate models are beneficial when the true objective function is expensive to evaluate, and a mechanistic model is unavailable--both common scenarios in complex bioprocess engineering systems.
\\
For example, fitting a simple mathematical relationship such as a polynomial or response surface to describe how fermentation yield varies with temperature is a straightforward application of surrogate modeling. Importantly, surrogates are deliberately less complex than the physical system they approximate. Rather than aiming for high-fidelity reproduction across the entire parameter space, they should provide sufficiently informative approximations to support informed decisions.
\\
A desirable property of a surrogate model is the ability to act as a universal approximator, meaning it can capture complex input–output relationships even when the true system behavior is unknown.
This capability is crucial in bioprocess engineering, where the underlying biological behavior is often unknown or only partially described. 
Returning to the fermentation example, the true relationship between yield and temperature may be highly nonlinear and poorly understood. Nonetheless, a well-chosen surrogate model with universal approximation capability can still capture the underlying trends and interactions effectively.
\\
While universal approximation is a desirable property, the ability to estimate predictive uncertainty is essential for \ac{bo}, and not all surrogate models offer this capability.
Common surrogate models that do provide uncertainty estimates include \acp{gp} probabilistic decision trees, and Bayesian neural networks. Among these, \acp{gp} are particularly popular in \ac{bo}. They provide uncertainty estimates alongside predictions and operate as non-parametric models that flexibly adapt to data without assuming fixed structural forms. Moreover, they perform well even with limited experimental data compared to other universal approximators such as neural networks. These characteristics make \acp{gp} well-suited for applications in bioprocess engineering.
In the following, we introduce \acp{gp} in more detail.

\subsubsection{Introduction to Gaussian Processes}
\label{sec:gps}
Suppose we want to model an unknown function $f(\mathbf{x})$, where $\mathbf{x}$ is the vector of design variables.
This function could be an enzyme's catalytic activity, which depends on temperature and pH.
A \ac{gp} defines a probability distribution over possible functions that could describe such a dependency, assigning
higher likelihood to functions that align with both prior knowledge of the system and observed data. Conceptually, a \ac{gp} extends the multivariate normal distribution to infinitely many variables, one for each possible input point in the domain, i.\,e.,~combinations of our design variables.
For a primer on the statistical background for \acp{gp}, such as multivariate normal distributions, conditioning, and marginalization, see Box~\ref{box:gp_maths}. Additionally, for further explanation on the link between multivariate normal distributions and \acp{gp}, we recommend the following tutorials~\cite{wangIntuitiveTutorialGaussian2023, gortlerVisualExplorationGaussian2019}.\\
Any finite set of function values drawn from a \ac{gp} follows a multivariate normal distribution.
Thanks to this property, \acp{gp} allow closed-form inference: we can compute predictions and uncertainties exactly using linear algebra.
This makes them computationally tractable, as the required calculations are efficient and do not involve approximations or iterative fitting.
\\
A \ac{gp} is fully defined by two components: the mean function, $m(\mathbf{x})$, which gives the expected function value at input $\mathbf{x}$ before observing data, and the covariance function, $k(\mathbf{x},\mathbf{x}^\prime)$, which quantifies the similarity of outputs between inputs $\mathbf{x}$ and $\mathbf{x}^\prime$. In practice, the covariance function determines the entries of the covariance matrix, $\boldsymbol{\Sigma}$.
\[
f(\mathbf{x}) \sim \mathcal{GP}(m(\mathbf{x}), k(\mathbf{x}, \mathbf{x}^\prime))
\]

\begin{figure}[ht!]
    \centering
    \includegraphics[width=0.68\textwidth]{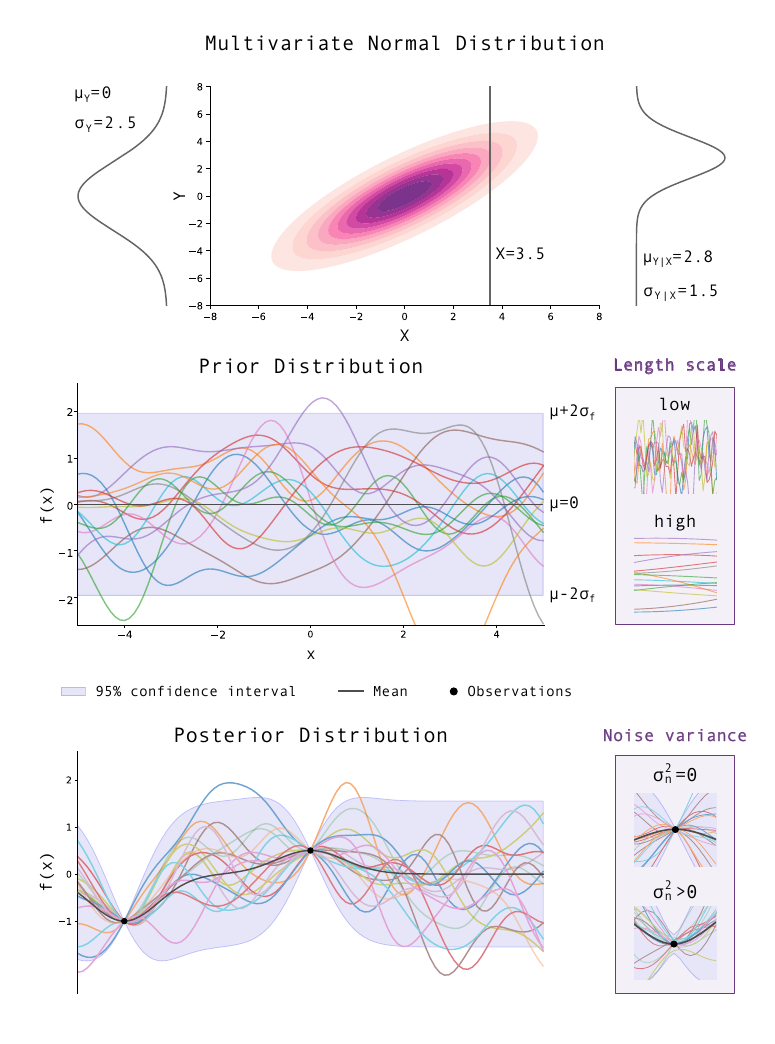}
    \caption{Overview of Gaussian processes:
(a) Marginal, joint, and conditional distributions of two correlated Gaussian random variables ($\mu_X = \mu_Y = 0$, $\sigma_X = \sigma_Y = 2.5$, $\rho = 0.8$). Left: marginal $p(Y)$ with $\mu_Y = 0$, $\sigma_Y = 2.5$. Center: joint contours with vertical slice at $X = 3.50$. Right: conditional $p(Y \mid X = 3.50)$ with $\mu_{Y|X} = 2.80$, $\sigma_{Y|X} = 1.50$.
The structure in (a) provides the foundation for \acp{gp}: while a multivariate normal distribution models joint behavior over a finite set of variables, a \ac{gp} extends this to an infinite collection of random variables, one for each input, such that any finite subset follows a multivariate normal distribution.
(b) Prior distribution: samples from a \ac{gp} prior with zero mean and a squared exponential kernel. The shaded region indicates the 95\% confidence interval. Inset: effect of kernel length scale $l$, where a lower $l$ yields more rapid variation and a higher $l$ produces smoother functions. 
(c) Posterior distribution: \ac{gp} posterior after conditioning on observations (black markers). Inset: higher noise variance $\sigma_n^2$ increases predictive uncertainty and flexibility of the posterior.
    }
    \label{fig:gp_concepts}
\end{figure}

\noindent Note that the term kernel is commonly used in place of covariance function in the literature; hereafter, we adopt the term kernel throughout this work.
\\
\noindent Initially, before observing any data, we define a prior distribution based on our assumptions (see Box~\ref{box:gp_maths}).
A crucial part of this prior is the choice of kernel, which encodes assumptions about properties such as smoothness, periodicity, or trends in the underlying function. 
One of the most popular kernels is the squared exponential (or radial basis function -- RBF) kernel, defined as
\[
k(\mathbf{x}, \mathbf{x}') = \sigma_f^2 \exp\left( -\frac{\lVert \mathbf{x} - \mathbf{x}' \rVert^2}{2\ell^2} \right) + \sigma_n^2 \, \delta(\mathbf{x}, \mathbf{x}'),
\]
where the hyperparameters are the length scale, $\ell$, signal variance, $\sigma_f^2$, and noise variance, $\sigma_n^2$.
The length scale, $\ell$, controls the rate at which correlations between function values decay with distance in the parameter space , affecting how much each observation informs predictions at nearby locations.
Intuitively, it determines how quickly the function is allowed to vary across the parameter space: small values lead to rapid changes over short distances, while large values enforce smoother, more slowly varying functions.
\\
This behavior is illustrated in Figure~\ref{fig:gp_concepts} (top inset), which shows the contrast between short and long length scales.
The signal variance, $\sigma_f^2$, sets the overall amplitude of the function, indicating how strongly it is expected to vary around its mean.
The noise variance, $\sigma_n^2$, captures uncertainty in the observations themselves, allowing the model to account for measurement noise. When $\sigma_n^2 = 0$, the model assumes noise-free data, whereas increasing this value allows for greater deviation between the modeled function and the observed data -- this effect is visualized in Figure~\ref{fig:gp_concepts} (bottom inset).

\noindent Finally, different kernels encode different assumptions about the underlying function. The squared exponential kernel implies smooth behavior, while the Matérn kernel introduces a smoothness parameter $\nu$ to model rougher functions, and periodic kernels are well-suited to capturing repeating trends. These are among the most commonly used kernels, though many others exist for specialized modeling needs. Section~\ref{sec:mip_cat} explores more advanced kernels.
\\
Once the choices for kernels and hyperparameters are made, the \ac{gp} can be used for \ac{bo}.
As we collect data, we use it to update our prior beliefs, forming a posterior distribution (see Box~\ref{box:gp_maths}). This updated \ac{gp} reflects both what we have observed and our remaining uncertainty about the function elsewhere.
The ability of \acp{gp} to perform such updates stems directly from the properties of the multivariate normal distributions discussed in Box~\ref{box:gp_maths}: marginalization allows prediction at new points, while conditioning lets us update our model as we observe new data.
\\
Thanks to the conditioning and marginalization properties of multivariate normals, \acp{gp} enable:

\begin{itemize}
\item \textbf{Prediction:} Estimating function values at new, unmeasured points using observed data.
\item \textbf{Quantifying Uncertainty:} Assigning confidence levels to predictions, often visualized by the confidence intervals.
\item \textbf{Continuous Refinement:} In \ac{bo}, new data points are chosen for observation to update and refine our model. How do we decide which experiment to run next? \textit{This is where acquisition functions come in!}
\end{itemize}
\noindent For readers eager to learn more about \acp{gp}, we recommend Gaussian Processes for Machine Learning by Rasmussen and Williams~\cite{rasmussenGaussianProcessesMachine2006}.
\clearpage

%% file: 02_Methodological_Background/4_opt_acquisition_fn.tex
\subsection{Acquisition Function}
\label{sec:acquisition}
\begin{figure}[htbp]
    \centering
    \includegraphics[width=\textwidth]{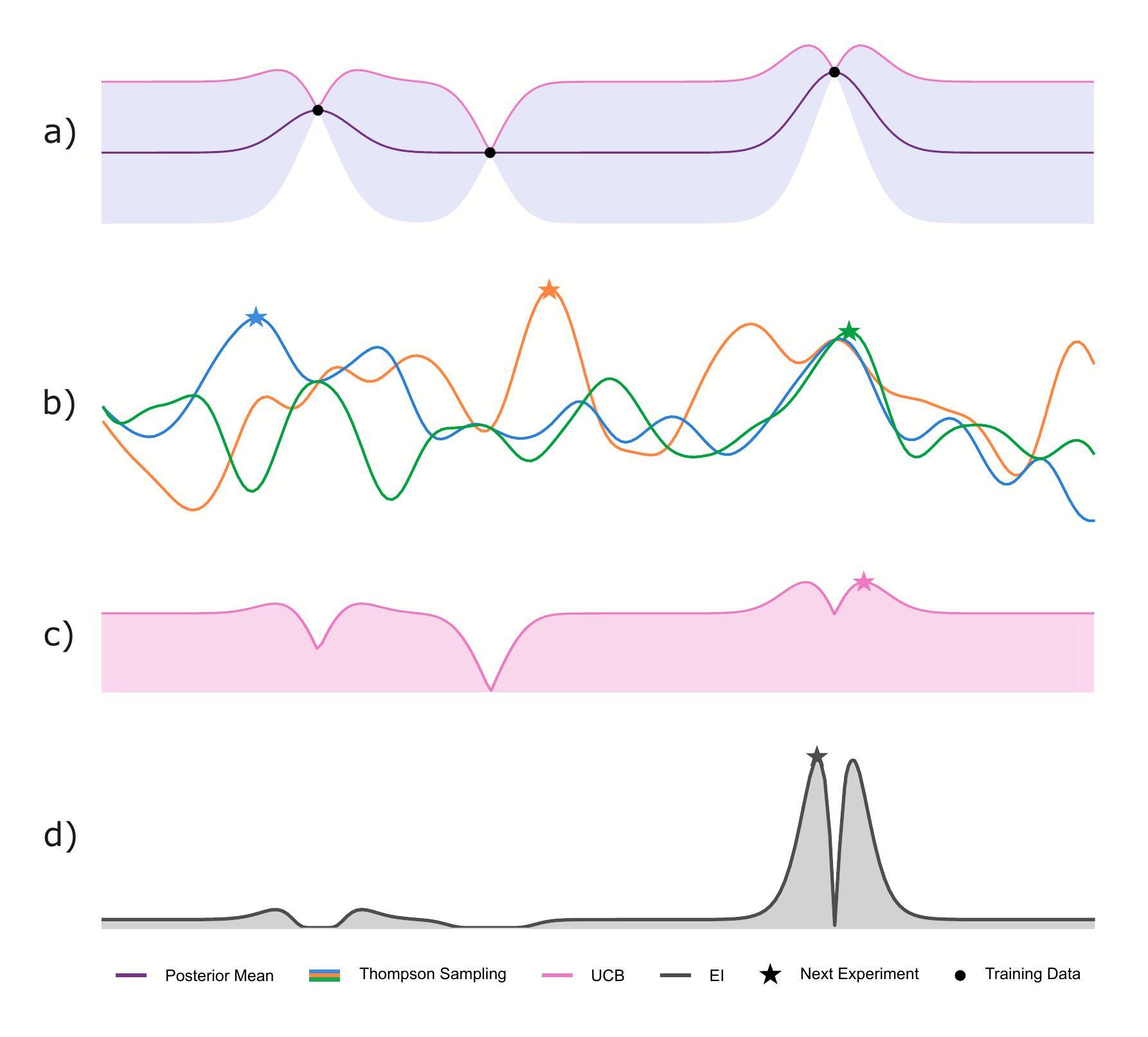}
    \caption{Comparison of \ac{bo} acquisition functions. a) The top row shows the \ac{gp} posterior with confidence intervals and observations. The right-most black circle represents the current highest observation -- which is the baseline for \ac{ei}. The solid pink line  visualizes the \ac{ucb}. The subsequent rows illustrate different acquisition strategies: b) Thompson sampling (sample functions from the GP posterior, shown in blue, green, and orange), b) \ac{ucb} (pink), and d) \ac{ei} (grey). Stars mark the next sampling points proposed by each acquisition function.}
    \label{fig:acquisition_functions}
\end{figure}

Iterative optimization typically proceeds via an optimization policy, i.\,e.,~a strategy for selecting the next observation(s) based on existing data.
In \ac{bo}, this policy is indirectly implemented through an \textit{acquisition function}, which assigns scores to candidate experiments and estimates their contribution to the optimization process, based on the current data, surrogate model, and optimization policy.
Most acquisition functions balance two key paradigms, \textit{exploration} and \textit{exploitation}.
Exploration refers to making observations, i.\,e., conducting an experiment, where predictive uncertainty is high, aiming to discover potentially promising regions in the parameter space.
In contrast, exploitation prioritizes regions with high predicted objective values, aiming to refine the current best estimate.
Effective optimization requires a careful balance of the exploration-exploitation trade-off. While many acquisition functions have been proposed, we focus on three widely used approaches: \ac{ei}, \ac{ucb}, and Thompson sampling (compare Figure~\ref{fig:acquisition_functions}):
\begin{itemize}
    \item \textbf{\Acf{ei}} quantifies the improvement expected from evaluating a candidate solution compared to the current best-known observation (gray area in Figure~\ref{fig:acquisition_functions}). It balances exploitation (sampling where the predicted mean is high) and exploration (sampling where uncertainty is large), guiding the search toward promising regions. Formally, \ac{ei} is defined as:
    \[
    \mathrm{EI}(\mathbf{x}; \xi) = (\mu(\mathbf{x}) - f(\mathbf{x}^\star) - \xi) \, \Phi\left( \frac{\mu(\mathbf{x}) - f(\mathbf{x}^\star) - \xi}{\sigma(\mathbf{x})} \right) + \sigma(\mathbf{x}) \, \phi\left( \frac{\mu(\mathbf{x}) - f(\mathbf{x}^\star) - \xi}{\sigma(\mathbf{x})} \right)
    \]
    Here, $\mu$ and $\sigma$ are the \ac{gp} predictive mean and standard deviation at point $\mathbf{x}$, $f(\mathbf{x}^\star)$ is the best observed value so far, and $\xi$ is a parameter that encourages exploration when set to a higher value. The term $\Phi(\cdot)$ denotes the cumulative distribution function of the normal distribution and captures the likelihood of improvement, while $\phi(\cdot)$, the corresponding probability density function, reflects the model's uncertainty. Together, these terms reward points that are either likely to outperform the current best or are uncertain enough to merit exploration. For more details on \ac{ei}, we refer to~\cite{garnettBayesianOptimization2023, agnihotriExploringBayesianOptimization2020}.
    
    \item \textbf{\Acf{ucb}} selects candidates by computing an upper confidence bound on the GP posterior mean, given by $\mu(x) + \gamma \sigma(x)$, where $\gamma$ controls the exploration--exploitation trade-off (a larger $\gamma$ places greater weight on uncertainty, promoting exploration). In Figure~\ref{fig:acquisition_functions}, $\gamma = 0.999$, indicating that the acquisition function strongly prioritizes uncertain regions. This upper bound appears as the pink curve overlaid on the \ac{gp} posterior, and the pink-shaded \ac{ucb} acquisition function peaks in areas where this bound is highest. The selected candidate (pink star) lies in a region of high uncertainty and moderate predicted value, exemplifying the behavior of \ac{ucb} in a high-$\gamma$ regime.
    \item \textbf{Thompson sampling} selects candidates by drawing a \textbf{random sample} of a function from the GP posterior (see Section~\ref{box:gp_maths}) and choosing the point that maximizes it. This approach captures both prediction and uncertainty in a single realization, naturally balancing exploration and exploitation without relying on explicit hyperparameters. In Figure~\ref{fig:acquisition_functions}, three sample functions are shown in blue, green, and orange, along with their respective optima, demonstrating the diversity in candidate selection across random samples.
\end{itemize}

\clearpage

%% file: 02_Methodological_Background/5_termination_criterion.tex
\subsection{Termination Criterion}
The final decision in each optimization loop, as shown in Figure~\ref{fig:bo-overview}, is whether to terminate the process or proceed with another set of experiments.
The termination criterion does not need to follow a specific mechanism and is often determined in advance by external decision makers.
For example, many projects are constrained by a limited number of experiments or a maximum allowable cost.
In other cases, termination may be based on achieving a specific optimization target, such as reaching a desired product titer or attaining a defined level of confidence in having identified the optimum.
Alternatively, the decision to terminate may be evaluated dynamically.
For instance, either an external agent (e.\,g., a human decision maker) or the algorithm itself may choose to stop based on the experimental results obtained so far.
In bioprocess engineering, the termination criterion should ideally be informed by experimental experts, who can often pre-allocate an experimental budget and advise on an appropriate target for the objective function, for example, a fraction of the maximum theoretical yield.

%% file: 03_Practical_Guidelines/1_case_study_context.tex
\section{Practical Guidelines for Bayesian Optimization in Bioprocess Engineering}
\label{sec:practical_guidelines}
With the methodological foundations established in the previous section, including the \ac{bo} workflow and its core elements such as surrogate models and acquisition functions, we now turn to practical implementation.
Although \acp{gp} and acquisition functions are now widely supported by open-source software libraries~\cite{balandatBoTorchFrameworkEfficient2020, durholtBoFireBayesianOptimization2024, fitznerBayBEBayesianBack2025, gardnerGPyTorchBlackboxMatrixMatrix2018, headScikitoptimizeScikitoptimize2021}, their application to the optimization of experimental systems presents additional challenges.
In bioprocess engineering, these challenges stem in part from the biological variability and measurement noise inherent to living systems. If not properly addressed, these factors can impair model accuracy and undermine the effectiveness of \ac{bo} workflows.
\\
In this section, we translate the general methodology of \ac{bo} with \acp{gp} into domain-specific best practices.
To contextualize these guidelines, we introduce a representative case study from biocatalysis: a crude cell extract of unknown composition that contains multiple enzymes.
A detailed characterization of the extract would be costly, but preliminary experiments indicate that it effectively converts a given substrate into a desired product without further processing.
Our objective is to maximize the reaction rate by optimizing a single design variable, the pH value.
No prior knowledge is assumed regarding the pH optima of the individual enzymes, making this a suitable black-box problem for illustrating \ac{bo} workflows.
We use this example throughout the following sections and conclude with an overview of the sequential optimization in Figure~\ref{fig:crude_extract_iterations}.

%% file: 03_Practical_Guidelines/2_experiments.tex
\subsection{Experiments}
Following the workflow outlined in Figure~\ref{fig:bo-overview}, a successful \ac{bo} study begins with careful consideration of the number of design variables, the strategy for selecting initial experiments, and how these choices affect the overall experimental budget. These early decisions shape the complexity of the parameter space and optimization problem, the structure of the initial dataset, and ultimately the representativeness of the surrogate model.

\subsubsection{Defining the Parameter Space}
Most bioprocess optimization studies are initialized from existing processes, which may involve laboratory-scale proof-of-concept experiments or previously established process configurations intended for improvement. In such settings, candidate design variables are typically selected based on biological hypotheses, prior experience, or practical constraints such as technical feasibility or material cost.
\\
While \ac{bo} is well-suited to handle experimental setups involving many design variables, it remains subject to the curse of dimensionality. Increasing the number of variables expands the parameter space, thereby exponentially increasing the number of experiments required for process optimization~\cite{shahriariTakingHumanOut2016, greenhillBayesianOptimizationAdaptive2020}. Users must therefore carefully balance the number of considered design variables with the available experimental resources from the outset.
\\
In classical \ac{doe} methodology, it is common practice to reduce the number of design variables before optimization in high-dimensional settings. Established sensitivity analysis and screening techniques, such as Plackett-Burman, fractional factorial, or one-factor-at-a-time, can help to identify and eliminate factors with negligible influence~\cite{cosenzaMultiinformationSourceBayesian2022, narayananAcceleratingCellCulture2025}. Translating this practice to \ac{bo} by reducing dimensionality prior to model fitting can, in principle, improve model fitting and evaluation, and alleviate visualization challenges~\cite{binoisSurveyHighdimensionalGaussian2021, greenhillBayesianOptimizationAdaptive2020, shanSurveyModelingOptimization2010, siedentopAvoidingReplicatesBiocatalysis2025}.  However, these approaches are typically local in nature and may fail to detect nonlinear effects or higher-order interactions.  As a result, any such reduction carries the inherent risk of overlooking interdependencies between variables, which may subsequently affect process optimization~\cite{schonlauScreeningInputVariables2006}. Given this limitation, dimensionality reduction should be applied with caution. If variables are excluded prior to \ac{bo}, a good practice is to re-evaluate them at the predicted optimum after \ac{bo} has been terminated. Naturally, this strategy cannot reveal optima located in entirely different regions of the parameter space.
Despite its prevalence in bioprocess engineering, pre-screening and design variable reduction may not be necessary for multi-dimensional \ac{bo} when the experimental budget is not limiting. Hvarfner et al.\ have recently demonstrated that traditional \ac{bo} performs well in high-dimensional settings, provided one can assume that the underlying system remains bounded in complexity as the dimensionality increases~\cite{hvarfnerVanillaBayesianOptimization2024}. They show that traditional \ac{bo} performs favorably in these cases to specialized high-dimensional \ac{bo} methods~\cite{erikssonScalableGlobalOptimization2019}. Guidelines for configuring \acp{gp} appropriately are discussed in Section~\ref{sssec:priors}.
\\
Once the design variables are selected, defining appropriate bounds for each parameter becomes a key step in shaping the optimization space. These bounds are generally informed by available expert knowledge, physical and biological limitations, or economic considerations. Substrate concentrations, for example, may be constrained by solubility limits or material costs. For our case study, this would mean choosing a wide pH range, as there is no prior knowledge regarding the pH optima of the enzymes present in the crude extract. To ensure that the optimum lies within the parameter space, we define bounds ranging from pH~3~to~11.

\subsubsection{Designing Initial Experiments}
The quality of the initial dataset strongly influences the representativeness of the fitted surrogate model~\cite{georgiouDeterministicGlobalOptimization2025}. Insufficient or poorly distributed initial coverage of the parameter space may leave large regions underexplored, which can degrade the quality of the surrogate model, cause unreliable uncertainty estimates, and ultimately misguide the acquisition function in proposing suboptimal experimental conditions~\cite{delucaComparisonStrategiesIterative2023, greifStructuredSamplingStrategies2025}. Space-filling designs such as Sobol sequences or Latin hypercube sampling are recommended to achieve adequate coverage of the parameter space. Their importance grows with dimensionality, as the risk of under-sampling regions increases rapidly with the number of design variables.
\\
In lower-dimensional settings, such as our one-dimensional pH optimization case study, equidistant grids supplemented by expert-guided manual experiments are typically a good choice to initialize the \ac{gp} model~\cite{savageHumanalgorithmCollaborativeBayesian2024}. Accordingly, we would perform five equidistant initial experiments in the specified pH range to ensure broad coverage of the parameter space. In addition, one data point near neutral pH would be included, a region that expert knowledge on enzyme activity suggests may be particularly promising.
\\
As a general rule of thumb, the spacing between initial points should roughly match the expected length scale of the \ac{gp} kernel. In practice, it is not uncommon to allocate approximately half of the total number of experiments to the initial dataset~\cite{narayananDesignBiopharmaceuticalFormulations2021, morschettFrameworkAcceleratedPhototrophic2017, siedentopAvoidingReplicatesBiocatalysis2025}.
\\
In scenarios where prior experimental data are available, for example, from preliminary screening experiments, these data points can be used for fitting the initial \ac{gp} model. However, augmenting such datasets with additional space-filling designs remains advisable to improve coverage of the parameter space. Further experiments can be systematically selected to fill sparsely sampled regions or generated using approaches like the expert-augmented \ac{doe} proposed by Savage and del Río Chanona~\cite{savageHumanalgorithmCollaborativeBayesian2024}.

\subsubsection{A Case Against Experimental Replicates}
Replicated experiments--such as triplicates--are commonly used to obtain precise pointwise estimates within the parameter space. In the context of \ac{bo}, however, such replication may represent a missed opportunity to extract maximal information about the response surface from each experiment~\cite{daiBatchBayesianOptimization2023, kreylingReplicateNotReplicate2018}. The inherent uncertainty quantification provided by \ac{gp} models allows single measurements to meaningfully inform both predictions and their associated uncertainties. This strategy remains robust even in the presence of occasional outliers~\cite{siedentopAvoidingReplicatesBiocatalysis2025}.
\\
Nonetheless, replicates remain valuable when used selectively to quantify systematic sources of uncertainty, such as batch effects. These considerations are explored in more detail in the discussion on hybrid modeling in Section~\ref{sec:challenges_experiment} and Box~\ref{box:hybrid_model}. In line with these considerations, we elect not to perform replicated measurements in the pH case study. As shown in Figure~\ref{fig:crude_extract_iterations}, each experimental condition is evaluated only once to maximize information gain.

%% file: 03_Practical_Guidelines/3_surrogate_model.tex
\subsection{Surrogate Model}
Once the initial dataset is established, the next critical step in the \ac{bo} workflow is fitting the \ac{gp} model. At this stage, users must make informed decisions about the model configuration to ensure that prior knowledge is appropriately incorporated, that the model yields robust predictions, and that it provides reliable uncertainty estimates to guide the acquisition function.
\\
Prominent open-source libraries for \ac{bo} and \ac{gp} modeling~\cite{balandatBoTorchFrameworkEfficient2020, durholtBoFireBayesianOptimization2024, fitznerBayBEBayesianBack2025, gardnerGPyTorchBlackboxMatrixMatrix2018, headScikitoptimizeScikitoptimize2021} offer reasonable default configurations for general-purpose applications. However, the success of \ac{bo} in bioprocess engineering relies on adapting these configurations to the specific characteristics and constraints of the domain. We illustrate best practices for this adaptation in general, as well as using the previously introduced case study on pH optimization.

\subsubsection{Mean Function}
In most bioprocess engineering scenarios, detailed prior knowledge of the underlying system behavior is limited. In such cases, we recommend using a zero-mean function and standardizing the observations prior to model fitting. This configuration provides a simple, uninformative prior that enables the \ac{gp} kernel to learn the underlying functional relationship directly from the data without introducing additional assumptions~\cite{vondeneichenControlParallelizedBioreactors2022, siedentopAvoidingReplicatesBiocatalysis2025, lapierreMulticycleHighthroughputGrowth2025}.
We follow this approach in the pH optimization case study.
\\
In contrast, scenarios that involve strong mechanistic prior knowledge, such as systems governed by Monod or Michaelis-Menten kinetics, may justify the use of non-zero mean functions. The same applies to systems that exhibit known asymmetries due to external constraints, such as a sharp decline in productivity beyond a specific temperature caused by enzyme denaturation. Parametric mean functions or constant offsets can be incorporated to capture these trends. Incorporating mechanistic insight into data-driven models such as \acp{gp} falls under the broader category of hybrid modeling. Further details on hybrid modeling approaches are provided in Box~\ref{box:hybrid_model}.

\subsubsection{Kernel}
In contrast to the mean function, selecting an appropriate kernel requires more nuanced consideration in bioprocess engineering applications. 
Smooth or locally periodic kernels may be suitable, depending on the experimental context and the anticipated behavior of the underlying process. When strong prior knowledge is available, specialized kernels can be employed to encode structural patterns in the response surface. For instance, locally periodic kernels are particularly suitable for processes exhibiting time-resolved cyclic dynamics~\cite{ortmannAutomatedInsulinDelivery2019}, such as pulsed substrate feeding in fed-batch fermentations. By embedding periodicity into the kernel, they enable more confident extrapolation even when only a few cycles are observed. Despite their potential, such kernels remain underutilized in bioprocess engineering, and implementation guidelines remain scarce.
\\
In most other cases, bioprocess engineering applications are well represented by smoothing kernels. The squared exponential kernel and the more general Matérn kernel family are particularly well suited. Their smoothing characteristics generally align well with typical expectations for a wide range of bioprocess responses~\cite{siedentopAvoidingReplicatesBiocatalysis2025, vondeneichenControlParallelizedBioreactors2022, rosaMaximizingMRNAVaccine2022}.
We therefore recommend adopting these kernels in the absence of strong prior knowledge that would favor more specialized alternatives. We recommend that practitioners evaluate multiple kernels during initial model fitting. Visualizing random sample functions drawn from the posterior distribution under different kernel choices provides valuable insight into how each kernel encodes prior assumptions about the behavior of the system. For example, sample functions from the squared exponential kernel typically exhibit very smooth, gently varying behavior, while Matérn kernels produce rougher sample functions. Comparing these qualitative behaviors with prior expectations about the system can guide the selection of an appropriate kernel, which in many bioprocess applications is often the squared exponential kernel. In our case study, since enzymatic activity is expected to vary smoothly with pH, we follow this rationale and select the squared exponential kernel.

\subsubsection{Hyperparameter Priors}
\label{sssec:priors}
\begin{figure}[ht!]
    \centering
    \includegraphics[width=1\linewidth]{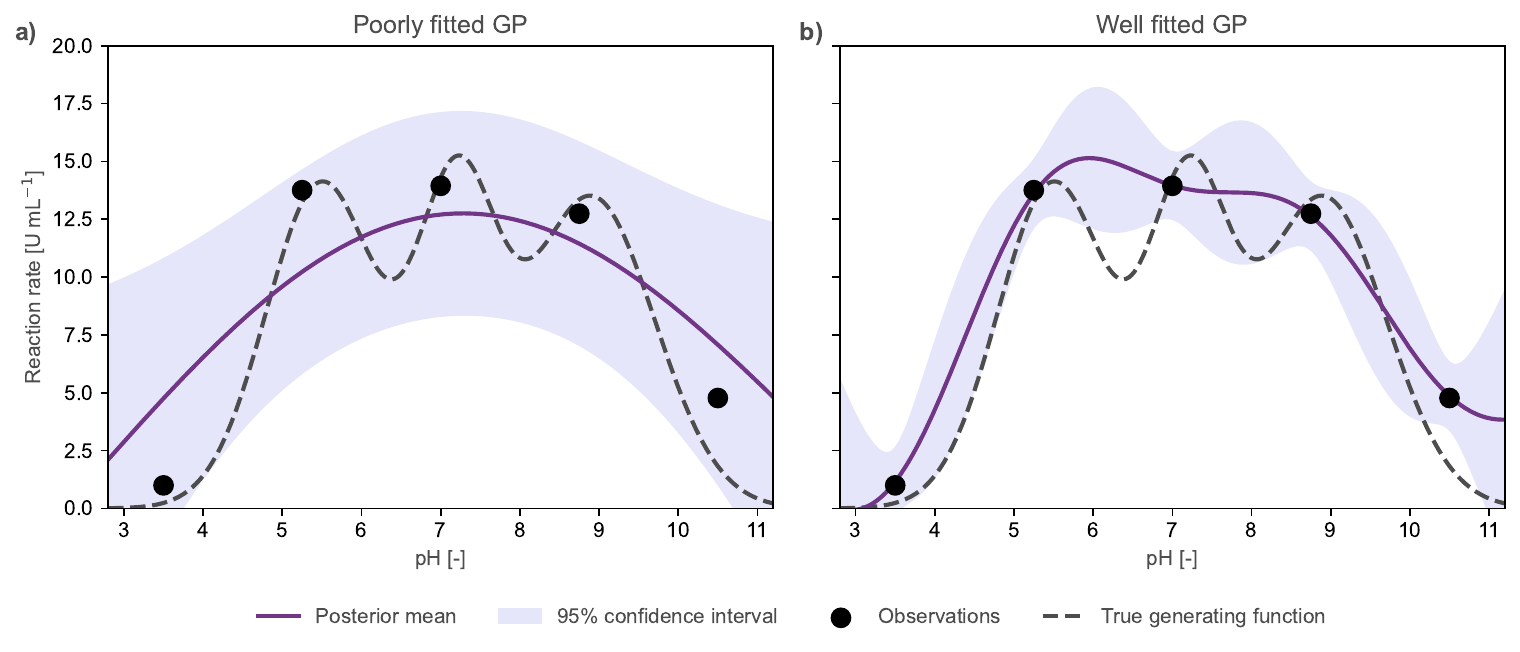}
    \caption{Comparison of well- and poorly-fitted \acp{gp}. Both panels display \ac{gp} models fitted to the five initial data points from the pH optimization case study. The true relationship between pH and reaction rate (dashed curve) remains unknown. Subplot a) illustrates the effect of inappropriate hyperparameter priors. Here, the length scale prior is centered at the full parameter space width, and the noise variance prior assumes \SI{50}{\percent} measurement noise. This configuration leads to underfitting. The model attributes most of the variance to noise, and the prediction uncertainty remains high even near the observations. In contrast, Subplot b) shows a well-fitted \ac{gp} that provides a plausible interpretation of the observations (black dots), attributing most of the observed variation to the underlying function. Here, we center the length scale prior at one-third of the parameter space and encode in the noise variance prior the belief that our observations carry \SI{10}{\percent} measurement noise. As a result, the predicted mean (violet curve) closely follows the data, and the uncertainty band narrows around the observed points. }
    \label{fig:fit_comparison}
\end{figure}

Once the mean function and kernel are selected, the critical task of configuring hyperparameter priors arises. For the widely used squared exponential kernel, these hyperparameters are the length scale~$\ell$, signal variance~$\sigma_f$, and noise variance $\sigma_n$ (Section~\ref{sec:acquisition}). As discussed in Box~\ref{box:gp_maths}, these hyperparameters are inferred following Bayes' theorem, which combines the specified priors with the information provided by the observed data. 
The choice of these priors can strongly influence the resulting posterior distribution, and thus the model fit~\cite{hvarfnerVanillaBayesianOptimization2024}, especially in low-data regimes.
\\
In practice, hyperparameters are typically estimated by maximizing the marginal likelihood of the \ac{gp}. However, the marginal likelihood surface is often highly non-convex and may exhibit multiple local optima~\cite{rasmussenGaussianProcessesMachine2006}. In low-data regimes, the problem becomes ill-posed, as numerous configurations can explain the data equally well. This ambiguity complicates the numerical optimization process. It is advisable to employ multi-start or other global optimization strategies for hyperparameter fitting to mitigate these issues.
\\
Figure~\ref{fig:fit_comparison} illustrates this challenge using our case study. Two distinct \ac{gp} fits are shown for the same five initial data points, differing only in their respective hyperparameter priors. Given that hyperparameter priors alone can determine whether a model is well-fitted, over-fitted, or under-fitted, we strongly advise practitioners to carefully assess their prior settings and inspect the resulting \ac{gp} fits. Visualizations of model predictions, uncertainty bands, and posterior samples are essential for diagnosing poor fits that may not be apparent from numerical statistics alone.
\\
To improve \ac{gp} model fits in scenarios with limited prior knowledge, we offer the following recommendations. For length scale priors, we recommend encoding how rapidly the underlying system behavior is expected to vary across the defined parameter space. A generally promising approach is to set the length scale in the range of one-quarter to one-half of the total parameter space width. As dimensionality grows, it is advisable to scale the length scale prior at a rate of $\ell \propto \sqrt{d}$, where $d$ is the number of design variables, provided this aligns with prior domain knowledge. Such an adjustment improves \ac{bo} performance in higher dimensions by offsetting the larger average distances between data points~\cite{hvarfnerVanillaBayesianOptimization2024}.
\\
Signal variance and noise variance often exhibit a trade-off during model fitting. The \ac{gp} can attribute observed variability either to underlying system behavior or to measurement noise~\cite{rasmussenGaussianProcessesMachine2006}. As illustrated in Figure~\ref{fig:fit_comparison}, poor prior specification can cause the model to misattribute this variability. To avoid such scenarios, we recommend avoiding informative priors for signal variance. In contrast, the prior for the noise variance should be informed by prior knowledge of the measurement system, such as precision estimates from previous calibration studies or historical experimental data.
\\
In the pH case study, we applied these recommendations by setting the prior for the length scale to one-third of the total parameter range. Additionally, we selected a prior for the noise variance based on an estimated measurement uncertainty of 5\%.
For the detailed implementation and an interactive Jupyter notebook to explore how different hyperparameter priors affect model fit, we refer to the accompanying Github repository~\cite{siskaLhelleckesBO_Empirical_ExamplesV0102025}.

\subsubsection{Isotropic and Anisotropic Length Scales}
In multi-dimensional design spaces, practitioners may choose between isotropic kernels, which use a single length scale shared across all design variables, and anisotropic kernels, often referred to as automatic relevance determination kernels, which assign separate length scales to each dimension. Isotropic kernels reflect the assumption that all design variables influence the underlying function at similar scales. In contrast, anisotropic kernels enable the model to capture variable-specific sensitivities and directional effects. Liang et al.\ report that anisotropic kernels tend to outperform isotropic ones when the design variables differ substantially in nature or scale~\cite{liangBenchmarkingPerformanceBayesian2021}. For example, consider extending the pH case study with a second design variable: temperature, which typically exhibits a single broad optimum. Using a single (isotropic) length scale for both pH and temperature could restrict the ability of the model to capture the sharper pH dependency illustrated in Figure~\ref{fig:crude_extract_iterations}. While this added flexibility of anisotropic kernels enhances model expressiveness, it also increases the risk of overfitting~\cite{greifStructuredSamplingStrategies2025}, especially in data-scarce scenarios.

%% file: 03_Practical_Guidelines/4_acquisition_function.tex
\subsection{Acquisition Function}
The acquisition function is the next core component in the \ac{bo} workflow. It proposes new experiments by leveraging both the predictive mean and uncertainty estimates of the \ac{gp} to balance exploration and exploitation. This section focuses on practical considerations for the commonly used acquisition function, as well as typical pitfalls when managing the exploration-exploitation trade-off. We restrict our discussion to sequential \ac{bo}, where one experiment is proposed at a time. Batch \ac{bo} strategies, which are highly relevant in bioprocess engineering for parallel experimentation, are discussed in Section~\ref{sec:advanced_bo}.

\subsubsection{Balancing Exploration and Exploitation}
\begin{figure}[ht!]
    \centering
    \includegraphics[width=0.95\linewidth]{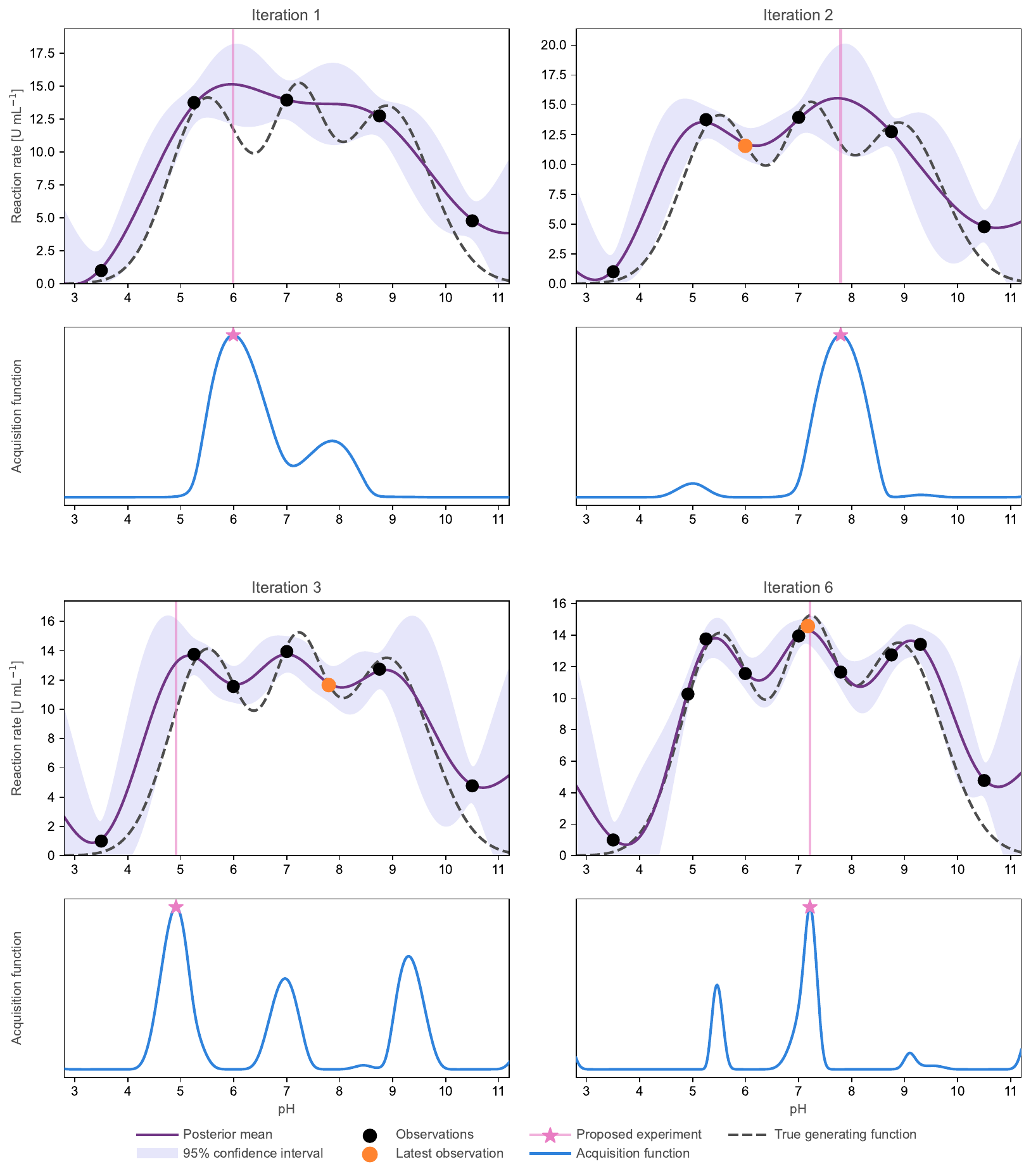}
    \caption{Progression of \ac{gp} model and acquisition function during \ac{bo} of the pH case study. Panels illustrate the evolution of the \ac{gp} model throughout multiple iterations of the \ac{bo} workflow. For each iteration, the upper panels display the \ac{gp} model fitted to the available observations. The data point proposed in the previous iteration is highlighted in orange. The lower panels show the corresponding \ac{ei} acquisition function. The pink line marks the location of the next proposed experiment, corresponding to the maximum of the \ac{ei}. No replicated experiments are performed. As the optimization proceeds, the \ac{gp} model incrementally improves its approximation of the system behavior, while the \ac{ei} acquisition function dynamically adjusts to balance exploration and exploitation.}
    \label{fig:crude_extract_iterations}
\end{figure}

\Acf{ei} and \acf{ucb} are among the most widely used acquisition functions in the literature. They are readily available in open-source libraries~\cite{balandatBoTorchFrameworkEfficient2020, durholtBoFireBayesianOptimization2024, gardnerGPyTorchBlackboxMatrixMatrix2018, headScikitoptimizeScikitoptimize2021}, easy to interpret, effective in low-dimensional problems, and often yield good performance with default settings. However, both rely on user-defined hyperparameters to adjust the trade-off between exploration and exploitation.
\\
Selecting these hyperparameters can substantially influence optimization outcomes, particularly during early iterations when the surrogate model is fitted on limited and potentially unrepresentative data. Overly exploitative strategies increase the risk of premature convergence to local optima. Georgiou et al.~\cite{georgiouDeterministicGlobalOptimization2025} report that exploitative acquisition functions are especially prone to failure when the initial dataset is sparse or clustered. This issue is particularly relevant in bioprocess engineering, where experiments are costly and the resulting data are often noisy and scarce. Nonetheless, \ac{ei} and \ac{ucb} remain well-suited options, provided that the initial data are of sufficient quality and hyperparameters are judiciously selected.
\\
Optimizing acquisition functions is non-trivial, as they are often highly non-convex and exhibit flat landscapes across large regions of the parameter space~\cite{binoisSurveyHighdimensionalGaussian2021}. These characteristics can lead to sub-optimal experiment proposals. Efforts to mitigate these issues include the development of logEI, a reformulation of \ac{ei} that improves numerical performance and has been shown to result in significantly improved optimization outcomes~\cite{amentUnexpectedImprovementsExpected2023}. Additionally, practitioners of \ac{bo} are advised to apply multi-start strategies and stochastic evolutionary algorithms to improve the robustness of acquisition function optimization~\cite{georgiouDeterministicGlobalOptimization2025, vincentImprovedHyperparameterOptimization2023}.
\\
Figure~\ref{fig:crude_extract_iterations} illustrates the application of \ac{ei} to the pH case study. Starting with the initial dataset in iteration~1, \ac{ei} selects exploratory experiments that expand coverage of the parameter space, thereby reducing predictive uncertainty. As model uncertainty decreases, \ac{ei} progressively shifts toward more exploitative proposals. By iteration~6, the algorithm converges to the global optimum, having effectively learned the shape of the objective function near its peak and identified the optimal pH for the crude extract at 7.2.

\subsubsection{Hyperparameter-Free Alternative}
Thompson sampling offers a practical alternative to traditional, hyperparameter-dependent acquisition functions. By randomly sampling functions from the \ac{gp} posterior, it naturally balances exploration and exploitation and fosters broader exploration of the design space. This stochasticity makes it particularly robust in early \ac{bo} iterations when data is sparse and uncertainty is high. 
More advanced information-theoretic acquisition functions, such as Predictive Entropy Search~\cite{hernandez-lobatoPredictiveEntropySearch2014}, share conceptual ground with Thompson sampling, as they also rely on sampling from the \ac{gp} posterior rather than using dedicated hyperparameters. These acquisition functions propose new experiments by estimating which would most reduce the uncertainty about the location of the global optimum. This property makes these acquisition functions especially well-suited for batch \ac{bo} (see Section~\ref{sec:advanced_bo}). However, their high computational cost has thus far limited their use in bioprocess engineering applications.
\\
Thompson sampling, too, can become computationally demanding in multi-dimensional settings, due to the sampling from the \ac{gp} posterior. In particular, drawing samples over a discretized grid in the parameter space becomes increasingly prohibitive as dimensionality grows, with the computational cost scaling cubically with the grid resolution~\cite{rasmussenGaussianProcessesMachine2006}.
Several strategies have been proposed to mitigate this burden, such as using more efficient sampling schemes or \ac{gp} approximations based on random Fourier features~\cite{rahimiRandomFeaturesLargescale2007, hernandez-lobatoPredictiveEntropySearch2014}. While such techniques can reduce runtime, they often compromise fidelity by failing to accurately capture the true \ac{gp} posterior in high-dimensional spaces~\cite{tonSpatialMappingGaussian2018, weselLargeScaleLearningFourier2021}. As experimental systems become more complex and parameter spaces expand, continued advances in scalable surrogate modeling and acquisition strategies will be essential for enabling practical and reliable \ac{bo}.

%% file: 04_Applications_Bioengineering/applications.tex
\clearpage
\section{Challenges of Applications in Bioprocess Engineering}
\label{sec:challenges_experiment}
While the previous section presented practical guidelines based on a base case, this section focuses on more specific, yet commonly encountered, challenges and additional considerations in the application of \ac{bo} to bioprocess engineering.
\subsection{Handling Experimental Biases in Bioprocess Engineering}
Following the general discussion, we now focus on several specific challenges arising from experimentation in bioprocess engineering.
In particular, we highlight \textit{batch effects}, \textit{positional biases}, and the \textit{calibration of measurement devices}.
\\
Batch effects refer to systematic biases between experimental batches or samples, caused by factors unrelated to the design variables~\cite{leekTacklingWidespreadCritical2010}.
Examples include variations in results when experiments are conducted on different days or by different experimenters.
In bioprocess engineering, batch effects are common due to the complexity of experimental systems, measurement technologies, and the inherent variability of living organisms~\cite{gohWhyBatchEffects2017, hanEvaluatingMinimizingBatch2022}.
Addressing batch effects in \ac{bo} requires extending the standard \acp{gp} framework, as illustrated by an empirical example in Box~\ref{box:hybrid_model}.
\\
Another common intricacy is positional bias, which frequently arises when working with microtiter plates~\cite{carausDetectingRemovingMultiplicative2017}.
These biases may result from temperature gradients~\cite{burtThermalCharacteristicsMicrotitre1979}, pipetting errors~\cite{pandyaStrategiesMinimizeVariability2010}, the fabrication of microtiter plates~\cite{krickaVariabilityAdsorptionProperties1980} or the plate reader (measurement device)~\cite{harrisonLocationDependentBiases1988}.
In classical \ac{doe}, batch effects and positional biases are typically addressed through randomization and blocking~\cite{montgomeryDesignAnalysisExperiments2020}, techniques that are also widely applied in bioprocess engineering~\cite{roselleMitigationMicrotiterPlate2016}.
However, these effects can alternatively be incorporated directly into the model.
Such model-based strategies are less commonly applied in \ac{bo} for bioprocess engineering, likely because most available software packages do not support them by default.
Nonetheless, recent studies such as von Eichen and Osthege et al.~\cite{vondeneichenControlParallelizedBioreactors2022} and Helleckes and Müller et al.~\cite{helleckesExploreExploitModelbased2023} provide useful examples.
In practice, it is worth investigating these experimental effects prior to applying \ac{bo}.
It is advised to either eliminate them to be able to use the basic practical guidelines outlined in Section~\ref{sec:practical_guidelines}, or to model them explicitly as illustrated in Box~\ref{box:hybrid_model} for batch effects.
\\
Finally, many objective functions in bioprocess engineering cannot be measured directly, for instance, titers inferred from fluorescence or absorbance assays.
In such situations, an accurate calibration of the measurement device is essential to convert the raw measurement signals into the desired target quantity.
This can be achieved with the help of calibration models, which in
practice are often applied before the converted data (e.\,g., an enzyme activity or a titer) is used in \ac{bo}.
In this context, the critical role of a well-suited calibration model should be emphasized.
Incorrect assumptions, such as linearity or neglecting a signal drift over time, can strongly bias the surrogate model and mislead the selection of subsequent experiments.
Using likelihood-based calibration models, experienced users can directly incorporate these assumptions into the surrogate model~\cite{helleckesBayesianCalibrationProcess2022}.
\\
These examples illustrate that the intricacies of bioprocess engineering require continuous fine-tuning of the surrogate model and tailored selection of priors and acquisition functions, as discussed further in Section~\ref{sec:advanced_bo}.
\subsection{Perspectives on User-Friendliness and Greater Adoption of Bayesian Optimization}
In contrast to traditional \ac{doe}, which typically follows a well-defined and structured design, setting up \ac{bo} involves a series of decisions that may be challenging for non-experts.
These include selecting the surrogate model, acquisition function, initial sampling strategy, and stopping criteria, along with potentially ongoing adjustments during implementation, making \ac{bo} more flexible but also more complex.
The development of bioprocess-specific frameworks or guidelines, such as this paper, could lower the entry barrier and facilitate a broader adoption within the field.
\\
Another major barrier is the limited availability of user-friendly software tools.
Whereas traditional \ac{doe} is supported by various commercial software packages with graphical user interfaces (e.\,g.~JMP by the SAS Institute or MODDE by Sartorius), most \ac{bo} applications in bioprocess engineering have been realized using software that requires intermediate to advanced programming and machine learning expertise.
Some studies have developed custom \ac{bo} software tools for bioprocess applications using MATLAB~\cite{freierFrameworkKrigingbasedIterative2016,freierMultiobjectiveGlobalOptimization2017} or Python~\cite{thompsonIntegratingTailoredRecurrent2023}.
However, most implementations rely on well-established general-purpose \ac{bo} libraries in Python.
The most frequently used libraries include:
\begin{itemize}
\item \texttt{BoTorch} and \texttt{GPyTorch} (e.\,g.,~\cite{claesBayesianCellTherapy2024, cosenzaMultiinformationSourceBayesian2022, cosenzaMultiobjectiveBayesianAlgorithm2023}),
\item \texttt{scikit-learn} and \texttt{scikit-optimize} (e.\,g.,~\cite{japelBayesianOptimizationUsing2022, pandiVersatileActiveLearning2022, narayananAcceleratingCellCulture2025}), and
\item \texttt{GPyOpt} (e.\,g.,~\cite{kandaRoboticSearchOptimal2022, yoshidaHighThroughputOptimization2023}).
\end{itemize}
In addition, two user-friendly packages have been recently developed in the chemical industry, namely \texttt{BoFire}~\cite{durholtBoFireBayesianOptimization2024}, driven by BASF, and \texttt{BayBe}~\cite{fitznerBayBEBayesianBack2025}, driven by Merck.
Although these packages offer extensive options for surrogate models and acquisition functions, thus greatly simplifying the development of \ac{bo} workflows, they still require intermediate programming skills.
Moreover, their documentation and application programming interfaces are often designed for the machine learning community, posing an additional challenge for users without that background.
\\
Finally, Bioprocess engineers often come from backgrounds in the natural sciences or biochemical engineering, which may not include training in \ac{bo} methods, the associated mathematics, or programming.
Targeted training programs can play an essential role in driving the adoption of these methods.
Emerging degree programs that combine biotechnology or natural sciences with computer science can improve such interdisciplinary competencies.
In addition, to support broader application of \ac{bo}, user-friendly tools and domain-relevant examples are crucial.
We hope that the guidelines provided in Section~\ref{sec:practical_guidelines} will contribute to the development of decision guidelines and their implementation in user-friendly \ac{bo} tools tailored to the needs of bioprocess engineering.

%% file: 05_Advanced_BO/advanced_bo.tex
\section{Advanced Bayesian Optimization for Bioprocess Engineering}
\label{sec:advanced_bo}

As illustrated in previous sections, \ac{bo} has emerged as a framework that can guide experimental design in biotechnology. After outlining general recommendations and current challenges for applying \ac{bo} in bioprocess engineering, we now turn to advanced \ac{bo} extensions. We believe that these extensions can be leveraged by domain scientists in the future to overcome the rigid, stage-by-stage paradigm of bioprocess development. Rather than optimizing strain selection, process conditions, and scale-up in isolation, holistic bioprocess development must acknowledge their interdependencies. \ac{bo} extensions, such as batch \ac{bo}, multi-fidelity \ac{bo}, and mixed-integer \ac{bo} can unify the traditional stages into one coherent optimization loop, paving the way for fully integrated bioprocess development.

\subsection{Batch Bayesian Optimization in the Era of High-Throughput}
Traditional \ac{bo} is a sequential strategy: at each iteration, a surrogate model (often a \ac{gp}) is updated using observations, and a single new experiment is selected by maximizing an acquisition function. This paradigm, however, does not adequately reflect the experimental realities of bioprocess engineering. In practice, experiment preparation (e.\,g., preculture, strain construction) and analytics (e.\,g., chromatography or mass spectrometry) often dominate timelines, yet can be performed in parallel. Running a batch of four experiments simultaneously, for example in parallelized bioreactor systems, requires no or only marginally more time than a single experiment, reducing the overhead significantly. Automated high-throughput platforms further enable parallel experimentation at scale. These experimental realities motivate the use of \textit{batch \ac{bo}}~\cite{daiBatchBayesianOptimization2023, gonzalezNewParadigmsExploiting2023}.
\\
Batch \ac{bo} extends the traditional framework by selecting a set of $q$ points (experiments) to evaluate in parallel at each iteration. 
Selecting a batch of $q$ points introduces additional complexity beyond standard \ac{bo}, as the points must collectively balance exploration and exploitation. Common strategies include:

\begin{itemize}
    \item \textit{Greedy or sequential conditioning:} Points are selected one at a time, each conditioned on the previous selections within the batch~\cite{gonzalezBatchBayesianOptimization2016}.
    
    \item \textit{Posterior sampling (Thompson sampling):} Points are selected by drawing random samples of functions from the \ac{gp} posterior and selecting their maximizers (see~\ref{sec:acquisition}). The inherent randomness of independent draws controls the diversity within the batch~\cite{kandasamyParallelisedBayesianOptimisation2018}. Determinantal point processes can extend on this concept by ensuring that selected points provide complementary information~\cite{navaDiversifiedSamplingBatched2022}.
    
    \item \textit{Joint acquisition maximization:} The acquisition function, whether it is \ac{ei}, \ac{ucb}, or predictive entropy search~\cite{hernandez-lobatoPredictiveEntropySearch2014}, is optimized jointly over the $q$ points. Though this approach can be the most principled, it can be computationally demanding to the point of intractability as $q$ increases~\cite{wilsonMaximizingAcquisitionFunctions2018}.
\end{itemize}

\noindent The integration of batch \ac{bo} into bioprocess engineering workflows represents a natural progression given the rise of high-throughput experimentation systems. Batch \ac{bo} provides a practical framework for leveraging high-throughput experimental capabilities, enabling faster, more data-efficient optimization in bioprocess development.
However, some modeling challenges continue to hinder its seamless adoption.
\\
Miniaturization and automated, reproducible experimental workflows expose systematic biases, such as batch and positional effects, as described in Section~\ref{sec:challenges_experiment}.
The prevalence of at-line measurement devices (e.\,g., photometers) and their associated assays demands calibration models. Addressing these complexities often requires dedicated hybrid models that extend the standard \ac{gp} framework with mechanistic components (see Box~\ref{box:hybrid_model}). These models must be iteratively refined in tandem with prior assumptions and experimental workflows, forming a time-consuming cycle that limits accelerated bioprocess development.
To unlock the full potential of (batch) \ac{bo} in high-throughput experimentation, the modeling community must shift from bespoke process models to shared libraries of reusable model components that capture systematic biases and establish formal knowledge-transfer practices (e.\,g., shared documentation, code repositories) to share methodologies and insights across processes. 

\subsection{Multi-Fidelity Bayesian Optimization for Bioprocess Scale-Up}

Experiments in bioprocess engineering can be conducted at varying levels of fidelity (accuracy of the experimental system).
For example, small-scale microtiter plates can rapidly generate large amounts of data at relatively low cost and low fidelity, while pilot-scale bioreactor runs provide higher-fidelity data that more accurately reflects industrial conditions; however, they are time-consuming, costly, and low-throughput.
Traditional \ac{bo} frameworks do not distinguish between such data sources.
In contrast, \ac{mfbo} is explicitly designed to integrate and balance information from these different levels of fidelity.
\\
Specifically, \ac{mfbo} models both the objective function (e.\,g., yield, productivity, titer) and the fidelity level of each data source. It treats experimental outcomes as being drawn from a hierarchy of fidelities, where:

\begin{itemize}
    \item \textit{Low-fidelity sources} (e.\,g., small-scale, simplified models, in silico simulations) are cheap and fast to query but provide only approximate insights into real-world performance.
    
    \item \textit{High-fidelity sources} (e.\,g., pilot plant runs, advanced simulations) are accurate but expensive and slow to obtain.
\end{itemize}

\noindent By jointly learning from both types of data, \ac{mfbo} can efficiently explore the parameter space, using low-fidelity data to guide the search while reserving high-fidelity evaluations for only the most promising candidates~\cite{savageMachineLearningassistedDiscovery2024}. Importantly, \ac{mfbo} can facilitate holistic bioprocess development across scales.
It has the potential to integrate traditionally sequential bioprocess development stages, namely the optimization of process conditions and scale-up, and unite small-scale screening with large-scale validation.

\subsubsection{Bioprocess Scale-Up}

Scaling bioprocesses involves navigating a landscape where conditions that perform well at small scales do not always translate directly to larger scales due to factors like mass transfer, heat transfer, and mixing limitations. \ac{mfbo} provides a structured, data-driven way to traverse scales efficiently by rapid exploration at small scales (e.\,g., microtiter plates, small bioreactors) to map broad trends and identify promising regions of the parameter space.
High-fidelity experiments are targeted only for conditions that appear optimal or highly informative (e.\,g., pilot-scale bioreactors), conserving valuable time and resources. At every iteration, \ac{mfbo} algorithms select not only the experimental conditions, but also the scale (fidelity) at which these experiments are to be run. 
When used for scale-up, \ac{mfbo} thereby balances cost and accuracy by ensuring that the budget and experimental effort are concentrated where they will have the greatest impact on process development.

\subsubsection{Multi-Fidelity Bayesian Optimization Algorithms}

\ac{mfbo} frameworks are designed to capture both the dependency between fidelities and between design variables~\cite{forresterMultifidelityOptimizationSurrogate2007}. Different \ac{mfbo} methods capture the fidelity dependencies differently. In general, their acquisition functions consist of different terms, including a penalty for cost (e.\,g., time, budget), another penalty for loss of information from using a lower (cheaper) fidelity from a higher (more expensive) one, and a final term for exploration-exploitation trade-off.
The latter is also used in traditional \ac{bo} to make decisions; in this case, however, the decisions comprise the fidelity as well as optimal experimental conditions.
It is also possible to combine batch and multi-fidelity \ac{bo} to create a framework that screens over multiple fidelities in high-throughput~\cite{folchCombiningMultifidelityModelling2023, mossGIBBONGeneralpurposeInformationBased2021}.
In this way, multi-fidelity \ac{bo} offers a framework for bridging the gap between small-scale screening and pilot-plant validation, supporting faster, more cost-effective bioprocess development.

\subsection{Handling Discrete Variables in Bayesian Optimization}\label{sec:mip_cat}
Bioprocess engineering usually entails the optimization of both biocatalyst and process conditions, for example the selection of an enzyme out of a candidate library while optimizing the reaction temperature and pH.
Most of the solutions discussed so far address the optimization of continuous design variables, while the selection of a catalyst is a discrete problem.
Such joined optimization problems are referred to as \textbf{mixed-integer problems}, which handle both discrete and continuous parameters.

\subsubsection{Pragmatic Approaches for Mixed-Integer Problems in Bioprocess Engineering}
The simplest way to handle mixed-integer problems is one-hot encoding, which entails creating binary dummy variables that encode which data belongs to which category in the discrete space.
However, this method is inefficient for larger spaces, since each potential category of the discrete variable (in our case, the catalyst) requires its own dummy variable that takes values of 0 or~1.
\\
In another naive approach, discrete variables can be treated as continuous variables and solutions are then rounded to the next integer to obtain the selected candidates~\cite{daultonBayesianOptimizationDiscrete2022}.
Over the past years, several other modifications of kernels for discrete variables have been proposed, such as transformations on the discrete parameters~\cite{garrido-merchanDealingCategoricalIntegervalued2020} or splitting continuous and discrete design variables in different kernels and using sums or products of both to combine them into a valid mutual kernel~\cite{ruBayesianOptimisationMultiple2020}.
Modern libraries such as BoTorch provide various options, including kernels that can capture the correlation of discrete variables, e.\,g.,~the intrinsic coregionalization model~\cite{bonillaMultitaskGaussianProcess2007}

\subsubsection{Advanced Methods for Encoding Biocatalyst Information}
Beyond the selection of discrete candidates, details about the biocatalyst, for example genetic information on a cell line or the three-dimensional structure of an enzyme, can be encoded into the kernel to enhance the learning process.
Deep kernel learning~\cite{wilsonDeepKernelLearning2016} extends \ac{bo} by combining the flexibility of deep neural networks with the uncertainty quantification of \acp{gp}.
While traditional \ac{bo} models with simple kernels work well in low-dimensional structured problems, they often struggle in complex, high-dimensional domains common in bioprocess engineering, such as the amino acid sequence of a protein, or molecular structures.
Deep kernel learning addresses this by transforming inputs through a neural network into a latent space where a standard \ac{gp} kernel can operate, enabling \ac{bo} to handle challenging parameter spaces more effectively \cite{gomez-bombarelliAutomaticChemicalDesign2018}.
\\
In bioprocess engineering, deep kernel learning shows promise in applications like protein engineering and synthetic biology.
For example, (variational) autoencoders, machine learning models that are specifically suited to encode complex information to an embedding or "fingerprint", can map protein sequences into a compact latent space that captures key properties like stability or binding affinity.
\ac{bo} can then efficiently search this space for optimal designs~\cite{romeroNavigatingProteinFitness2013}.
These techniques of latent \ac{bo} are already extensively used in the realm of chemical engineering (e.\,g.,~\cite{mossBOSSBayesianOptimization2020,griffithsConstrainedBayesianOptimization2020}), thus showing potential for the encoding of biological properties in future applications.
\\
Beyond deep kernel learning, neural networks following the transformer architecture are of interest.
Transformers became famous in large language models such as ChatGPT, and are increasingly applied in the era of protein language models to encode protein sequences.
The most famous example is AlphaFold~\cite{jumperHighlyAccurateProtein2021}, which can be used to predict protein folding from sequence data.
Similar to autoencoders, protein language models are efficient in generating embeddings out of protein sequences.
In addition, graph neural networks are frequently applied to represent small molecules as molecular fingerprints~\cite{duvenaudConvolutionalNetworksGraphs2015, kwonExploringOptimalReaction2022}.
Protein language models and graph neural networks thus represent further options to encode biocatalyst information in the \ac{bo} workflow for future applications.

\subsection{Transfer Learning for Bayesian Optimization in Bioprocess Development}
In bioprocess engineering, researchers often work with families of related systems--such as optimizing media formulations across different microbial strains or fine-tuning enzyme activity across a panel of homologous proteins. In these scenarios, data from past experiments on related systems can provide valuable information to accelerate optimization of a new system. 
\\
\textit{Transfer learning} for \ac{bo} offers a way to accelerate experimental campaigns by leveraging knowledge from related tasks. 
\\
A key approach for transfer learning in \ac{bo} is the use of multi-output \acp{gp}, also known as multi-task \acp{gp} or co-kriging. Multi-output \acp{gp} model correlations not only between inputs and outputs but also across different, related tasks (e.\,g., strains, scales, or enzyme variants). By capturing these relationships, multi-output \acp{gp} allow prior data to improve predictions and guide the selection of new experiments for the current task. 
\\
Typical advantages of transfer learning in bioprocess development include:

\begin{itemize}
    \item \textit{Leveraging historical data:} make the most of prior experimental campaigns by reutilizing them in new tasks.
    
    \item \textit{Supporting generalization across related systems:} facilitates process development when moving between strains, enzymes, or scales.
\end{itemize}

\noindent Interestingly, transfer learning and multi-fidelity are closely related, as multi-fidelity can be seen as a particular case of transfer learning where the learning is transferred across scales.
Early case studies explore options for transfer learning in \ac{bo} for bioprocess engineering, for example work by Hutter et al.~\cite{hutterKnowledgeTransferCell2021}, who explored the encoding of different pharmaceutical products with embeddings, or Helleckes et al.~\cite{helleckesNovelCalibrationDesign2024}, who investigated meta learning of prior distributions for \ac{gp} parameters across cell lines.
However, most advanced options of transfer learning between processes and different biocatalysts remain underexplored and pose an interesting challenge to the machine learning community.

\subsection{Incorporating Models as Mean Functions in Bayesian Optimization}
Similarly to transfer learning, further incorporation of domain knowledge can enhance the traditional \ac{bo} workflow, for example, through the use of empirical models as mean functions or the application of hybrid models as surrogates.
In many bioprocess engineering applications, prior knowledge about a process is available in the form of mechanistic models, empirical correlations, or simulation frameworks (e.\,g., kinetic models, bioreactor dynamics, pathway simulations). A natural extension of \ac{bo} is to incorporate such models directly into the surrogate by using them as the \textit{mean function} of the \acp{gp}.
\\
In this setup, if $m(\mathbf{x})$ is the output of a mechanistic or empirical model at input $\mathbf{x}$, the \ac{gp} is applied to the residual $r$:

\[
y(\mathbf{x}) = m(\mathbf{x}) + r(\mathbf{x}), \quad r(\mathbf{x}) \sim \mathcal{GP}(0, k(\mathbf{x}, \mathbf{x}'))
\]

\noindent While not yet standard in off-the-shelf \ac{bo} frameworks, the integration of models as mean functions is an easy way to combine domain knowledge with data-driven design in bioprocess engineering.
\\
Beyond the use of mechanistic models as \acp{gp} mean functions, more complicated hybrid model structures are available.
Here, the \acp{gp} is replaced altogether by a probabilistic model.
Such an extension is discussed in Box~\ref{box:hybrid_model}.
\\
Other extensions to \ac{bo} relevant in bioprocess engineering include preferential \ac{bo} (using user preference over different options A and B)~\cite{gonzalezPreferentialBayesianOptimization2017, coutinhoHumanintheloopControllerTuning2024}, human-in-the-loop (using human expert feedback in the decision making) \ac{bo}~\cite{savageMachineLearningassistedDiscovery2024, biswasDynamicBayesianOptimized2024}, high-dimensional \ac{bo} (using a high number of design variables)~\cite{erikssonScalableGlobalOptimization2019,tangTRBEACONSheddingLight2024}, multi-objective (using more than one optimization target) \ac{bo}~\cite{hernandez-lobatoPredictiveEntropySearch2016}, and safe \ac{bo} (using safety constraints, for example avoiding unsafe experimental setups) ~\cite{petsagkourakisSafeRealTimeOptimization2021,paulsonAdversariallyRobustBayesian2022}.

%% file: 06_Reproducibility_and_Data_Deposition/data_deposition.tex
\section{Reproducibility and Data Deposition}
\label{sec:reproducibility}
\ac{bo} operates at the interface between wet-lab experimentation and computational modeling. As such, rigorous data organization and metadata collection are required across both domains to ensure full reproducibility and scientific validity of experimental and analytical workflows, including research software.

\subsection{Considerations for Experimental Workflows}
In bioprocess engineering, experimental workflows are often highly diverse and tailored to specific processes, posing particular challenges for standardized metadata collection. For instance, consider manually conducted biocatalysis experiments involving enzymes that are produced and purified in-house, where each measurement is costly, time-consuming, and yields very limited data. In such scenarios, reproducibility depends on meticulous record-keeping. Essential information includes enzyme source and purification protocols, preparation and storage conditions of stock solutions, and operational parameters such as temperature profiles and time delays during batch processing~\cite{tiptonStandardsReportingEnzyme2014, malzacherSTRENDABiocatalysisGuidelines2024}. Fully automated laboratory platforms ease reproducibility through scripted protocols and consistent experiment execution. To fully realize these benefits, users must also manage the additional layers of metadata, including experiment scripts, proprietary software configurations, and detailed version control of hardware and software components~\cite{jessop-fabreImprovingReproducibilitySynthetic2019}. 
\\
Recent advances in metadata standards promote comprehensive contextual documentation. Initiatives such as EnzymeML~\cite{lauterbachEnzymeMLSeamlessData2023}, STRENDA~\cite{tiptonStandardsReportingEnzyme2014, malzacherSTRENDABiocatalysisGuidelines2024}, and SBML~\cite{huckaSystemsBiologyMarkup2003, keatingSBMLLevel32020} provide structured frameworks for data exchange and continue to evolve in response to the increasing complexity of modern laboratory practice. These standards capture essential experimental details, such as reaction vessel types (e.\,g., microtiter plates, shake flasks), reaction conditions (e.\,g., temperature, agitation speed, substrate concentrations), and procedural steps. However, as automated laboratory platforms become increasingly prevalent, existing standards have struggled to keep pace with the complexity and volume of metadata they generate. To bridge this gap, Mione et al. have recently proposed a labeled property graph-based framework that integrates with laboratory workflow management systems to enable the automated collection of both experimental and its associated metadata~\cite{mionePropertyGraphSchema2025}. A key recommendation is to proactively record rich, machine-readable metadata that goes beyond the current requirements of prevailing standards. Once uniquely stored, this information can later be reused and integrated into emerging standards via automated conversion tools. Exhaustive data capture should be considered an investment in long-term reproducibility.

\subsection{Considerations for Computational Workflows}
Proper version control and documentation are essential to ensure that algorithmic decisions remain traceable over time. According to the FAIR principles, both data and software should be findable, accessible, interoperable, and reusable~\cite{wilkinsonFAIRGuidingPrinciples2016}. Importantly, these requirements also apply in proprietary contexts, even when the data or software are not open or freely available.
\\
On the computational side, the open-source software community routinely employs platforms such as GitLab or GitHub to support sustainable software development and reproducible computational science. In \ac{bo} workflows, git-based version control should likewise track every script, configuration file, and structured data reference, from initial experiment designs through hyperparameter choices, to guarantee full traceability of decision policies and algorithmic performance~\cite{sandveTenSimpleRules2013}. This is especially critical for \acp{gp}, where choices of length scale and noise variance priors often undergo revision between \ac{bo} iterations. 
\\
Hybrid process models in \ac{bo} (e.\,g., Box~\ref{box:hybrid_model}) further underscore the necessity to integrate computational and experimental metadata. Details such as well positions in microtiter plates or batch identifiers for cryogenic cultures must be versioned alongside model configurations to enable detection of systematic biases, including batch or positional effects~\cite{helleckesExploreExploitModelbased2023}. Electronic laboratory notebooks can streamline this by both documenting and aggregating essential data in machine-readable formats~\cite{dunieImportanceResearchData2017, potthoffProceduresSystematicCapture2019}. During hybrid process model development, the iterative feedback between modeling and experimental workflow drives updates to both. All of these human-in-the-loop decisions must be captured.
\\
Importantly, metadata must capture every intervention in the \ac{bo} workflow, including override or adaptation of algorithmic output, to provide a comprehensive, transparent history of input data, evolving model configurations, and corresponding outputs across successive \ac{bo} iterations. A holistic approach not only enhances reproducibility but also facilitates retrospective analysis, rigorous model validation, and reproducible evaluation of hyperparameter priors and their influence on model performance. Ultimately, thorough documentation and systematic data deposition are indispensable for ensuring the scientific robustness and translational potential of \ac{bo}, not only in bioprocess engineering.

%% file: 07_Conclusions/conclusions.tex
\section{Concluding Remarks}
\label{sec:conclusion}
\ac{bo} has garnered interest from scientists across disciplines, not only for optimizing experimental setups, reactions, and materials, but also as a tool for hyperparameter tuning in the context of machine learning workflows.
In bioprocess engineering, real-world applications of \ac{bo} remain nascent, posing several inspiring challenges for the coming years.
\\
From a technical perspective, more experience is required to identify demands and define standards for selecting surrogate models, prior distributions, and suitable acquisition functions.
Additionally, some optimization problems require customized algorithms capable of navigating parameter spaces with complex design variables, such as genetic information.
While this review presents recommendations and best practices for bioprocess engineering, we hope that these will be tested, evaluated, and further developed by domain scientists in the upcoming years.
At the same time, it aims to guide machine learning practitioners in understanding the unique demands of bioprocess engineering and in developing new algorithms and methods tailored to those challenges.
\\
More broadly, overcoming traditional experimental paradigms--such as triplicate designs, which are neither required nor beneficial for \ac{bo}--remains a practical challenge.
Nonetheless, rethinking experimental designs is essential to unlock the full potential of \ac{bo}, as demonstrated by early pioneering applications.
In addition, current developments toward FAIR data and software documentation, all increasingly discussed in the context of artificial intelligence, will greatly facilitate the use of \ac{bo}.
\\
Overall, we foresee \ac{bo} evolving from a technically demanding method applied by specialists and in isolated case studies to a standard tool for experimental design in areas such as biocatalysis, strain screening, and bioreactor development.
Achieving this transition will require both empowering and sensitizing domain scientists, as well as familiarizing machine learning engineers with the unique challenges of bioprocess engineering and experimental design.
As such, it is important for the field to develop best practices for addressing the specific intricacies of biological systems and their associated measurement setups in the coming years.

%% file: 08_Boxes/background_gps.tex
\newpage
\section{Box: Mathematical primer for Gaussian processes}
\label{box:gp_maths}
\subsection{Normal (Gaussian) Distributions}
To understand \acp{gp} intuitively, it is helpful to begin with its fundamental building block: the normal distribution. A real-valued random variable \(X\) follows a normal (Gaussian) distribution if:
\[
X \sim \mathcal N(\mu,\sigma^{2}),
\quad
p(x)=\frac{1}{\sqrt{2\pi\sigma^{2}}}\,
      \exp\!\Bigl[-\tfrac{(x -\mu)^{2}}{2\sigma^{2}}\Bigr],
\]
where the mean, \(\mu\), and the variance \(\sigma^{2}>0\) fully characterize the distribution.
Here, \(p(x)\) is the probability density function of the normal distribution.
\\
In practice, we are often interested in modeling more than one random variable.
This leads us to the multivariate normal distribution, or ``joint" normal distribution -- a generalization of the univariate normal to higher-dimensional spaces.
A vector of random variables, $\mathbf{X} = (X_1,\ldots, X_k)^\top$, follows a multivariate normal distribution if every subset of variables, or any weighted combination of them, is also normally distributed.
\\
We write this multivariate normal distribution as:
\[
\mathbf X \sim \mathcal N(\boldsymbol\mu,\Sigma),
\]
Here, \(\boldsymbol\mu\) is the mean vector, and \(\Sigma\) is a symmetric, positive-definite covariance matrix. Each entry \(\Sigma_{ij}\) quantifies how much $\mathbf{x}^i$ and $\mathbf{x}^j$ tend to increase or decrease together, i.\,e., how they co-vary.
Two properties make this distribution especially powerful:

\begin{itemize}
    \item \textbf{Marginalization} -- Any subset of variables from a multivariate normal distribution is itself normally distributed.
    \item \textbf{Conditioning} -- Observing some variables allows us to update our beliefs about the others, based on the known correlation between them, yielding a conditional distribution that is also multivariate normal.
\end{itemize}

\subsection{Bayes' theorem}

\acp{gp} are Bayesian models, meaning they rely not only on observed data but also incorporate prior beliefs through the kernel and mean function, which together define a prior distribution over possible objective functions in \ac{bo}.
Bayes' theorem provides the governing framework for updating the prior beliefs in light of new data that becomes available through experiments.
Let $\theta$ denote the model parameters, and $D$ the observed data. Bayes' theorem is given by:

\[
    p(\theta \mid D) = \frac{p(\theta) \cdot p(D \mid \theta)}{p(D)}
\]

\noindent Here, $p(\theta)$ is the prior distribution, i.\,e.,~the probability distribution we define for the parameters without having observed any data.
$p(D \mid \theta)$ is the likelihood, which expresses how likely the observed data is under a given choice of parameters.
The denominator $p(D)$, known as the evidence, serves as a normalization factor.
For continuous variables, it is given by the integral $\int p(\theta) \cdot p(D \mid \theta) \, d \theta$.
The result, $p(\theta \mid D)$, is the posterior distribution, reflecting our updated beliefs after observing the data.
\\
In \acp{gp}, these principles are applied when inferring the hyperparameters of the mean and kernel functions, such as the length scale and noise variance (see Section~\ref{sec:gps}).
These hyperparameters can be treated as random variables and updated using observed data, resulting in a posterior distribution over plausible hyperparameter values.
While full Bayesian inference is supported in  probabilistic programming frameworks such as PyMC~\cite{abril-plaPyMCModernComprehensive2023}, most widely used \ac{gp} and \ac{bo} libraries (e.g.,~\cite{balandatBoTorchFrameworkEfficient2020, gardnerGPyTorchBlackboxMatrixMatrix2018, headScikitoptimizeScikitoptimize2021}) rely on point estimates via marginal likelihood maximization instead.
The computational challenges of this approximation are discussed in Section~\ref{sssec:priors}.
The choice of prior functions for \acp{gp} will be further discussed in Sections~\ref{sec:practical_guidelines} and~\ref{sec:advanced_bo}.

%% file: 08_Boxes/hybrid_model_pymc.tex
\section{Box: Extension to Other Probabilistic Models as Surrogates for BO}
\label{box:hybrid_model}
\begin{figure}[ht!]
    \centering
    \includegraphics[width=\textwidth]{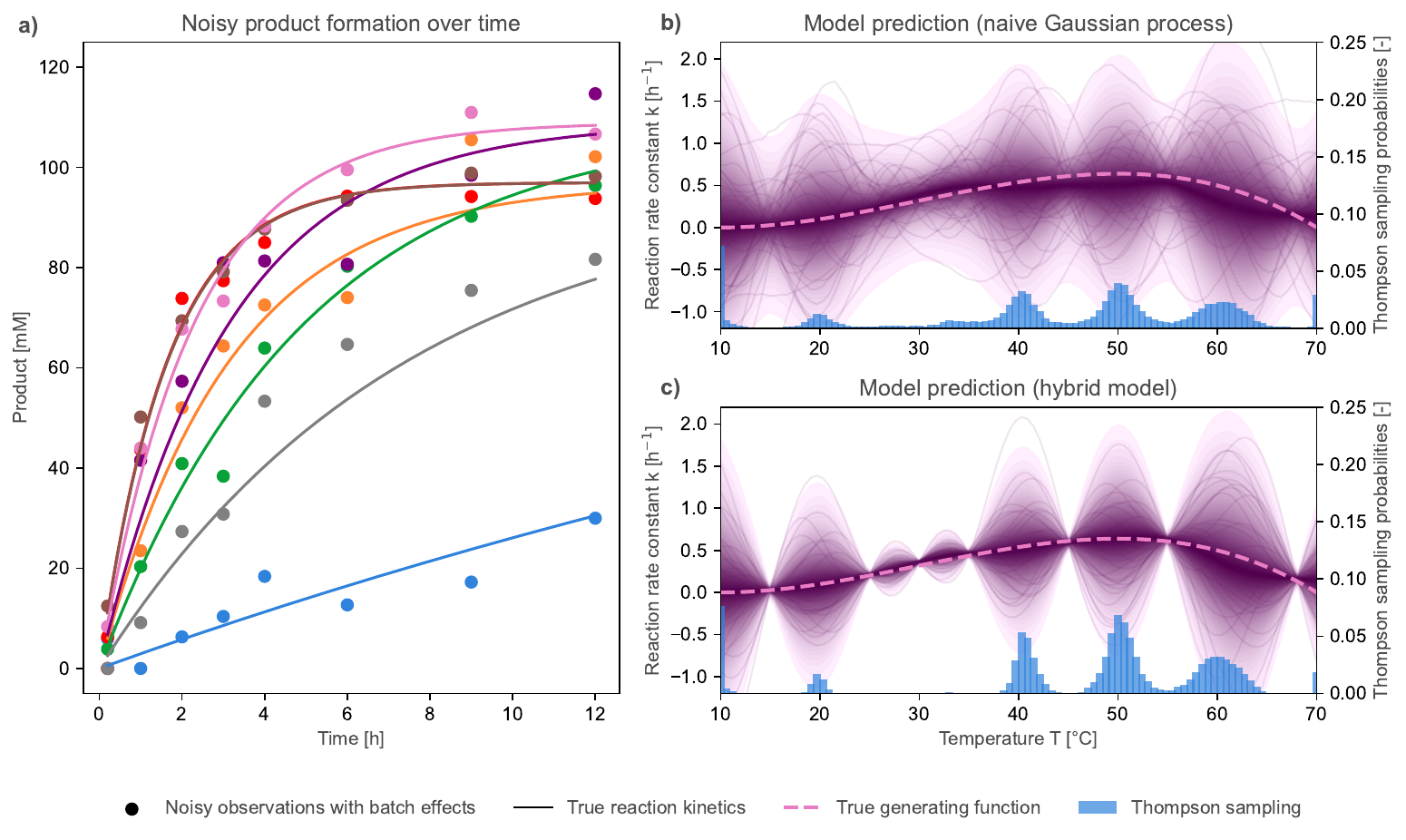}
    \caption{Biocatalysis case study with batch effects, comparing a naive \ac{gp} and a hybrid model as surrogates.\\
    This case study describes the optimization of the reaction temperature of an uncharacterized enzyme. The time series data of product formation for a single reaction at different temperatures can be measured (left). Experiments are conducted batch-wise in shake flasks (four experiments with different designs per batch), between which the substrate concentration changes due to an undesired batch effect. This example demonstrates more complex and thus realistic experimental conditions. In the left panel, the noisy observation of product concentrations over time can be seen, influenced by the respective batch effect. The following temperatures were used to generate in silico datasets: \SI{15}{\celsius} (blue), \SI{25}{\celsius} (green), \SI{30}{\celsius} (orange, purple), \SI{35}{\celsius} (pink), \SI{45}{\celsius} (red), \SI{55}{\celsius} (brown), \SI{68}{\celsius} (gray).
    The right two panels compare the fitted surrogate model (purple), as well as the distribution of probability when repeatedly applying Thompson sampling as optimization policy on the dataset of the current iteration of \ac{bo} (blue bars). The purple lines visualize random samples from the \acp{gp}, thus indicating the uncertainty around the mean prediction. A \ac{gp} hybrid model that is taking a random batch effect and the first-order mass action into account (bottom) captures the true relationship (pink dashed line) more efficiently than a naive \ac{gp} model (top). The latter was trained on reaction rates that were extracted from the time series data (left panel) via regression, which is a standard method in biocatalysis.}
    \label{fig:hybrid_model_pymc}
\end{figure}

\Acp{gp} remain the most frequently used surrogate models for \ac{bo} due to their favorable properties described in Section~\ref{sec:gps}.
However, bioprocess engineering case studies often come with prior knowledge on biological parameters, for example reaction kinetics.
As discussed in Section~\ref{sec:advanced_bo}, such cases may motivate the use of mechanistic models as mean functions for the \ac{gp}, or a shift to another probabilistic surrogate model altogether.
\\
In this box, we show an example of a simulated case study from biocatalysis that warrants sufficient prior knowledge to motivate the use of hybrid models (Figure~\ref{fig:hybrid_model_pymc}).
The case study describes the optimization of the reaction temperature of an enzyme, which follows first-order reaction kinetics.
We assume that observations of product formation for a single reaction at different temperatures can be measured (Figure~\ref{fig:hybrid_model_pymc}, left).
Experiments are conducted batch-wise in shake flasks (four experiments with different designs per batch), between which the substrate concentration changes due to an undesired batch effect.
This can be seen in the deviations between the measured data (dots) and the noise-free reaction curve (solid lines) in Figure~\ref{fig:hybrid_model_pymc} (left panel).
This example demonstrates more complex experimental conditions and requires the use of \ac{bo} extensions.
\\
We compare two approaches: a) the use of a naive \ac{gp} and b) a hybrid model, which takes a \ac{gp} in combination with mechanistic equations for reaction kinetics.
Using the probabilistic programming framework PyMC~\cite{abril-plaPyMCModernComprehensive2023}, we set up a model that describes the first-order reaction kinetics of the enzyme.
For product formation, we can explicitly formulate the reaction as:
\begin{equation}
    P_t = S_0 \cdot \left(1 - e^{(-k \cdot t)}\right)
    \label{eq:product_form}
\end{equation}

\noindent with $P_t$ as the product concentration at time $t$, $S_0$ as the initial substrate concentration and $k$ as the reaction rate constant.
\\
As shown in the plotted data (Figure~\ref{fig:hybrid_model_pymc}, left), the reaction rate constant $k$ is temperature-dependent, i.\,e., becoming a function $k(T)$.
Moreover, batch-wise experimentation usually implies that reaction media, biocatalyst, and the experimental setup are subjected to small deviations, such as fluctuation of the temperature, inexact processing times, or manual pipetting errors.
Here, we simulated small deviations in the initial substrate $S_0$ to mimic such a batch effect.
\\
The hybrid model uses a latent \ac{gp} (i.\,e., one where the predicted outcome cannot directly be measured) to describe the relationship between reaction rate $k$ and the temperature $T$:
\begin{equation}
k(T) \sim \mathcal{GP}(m(T), k(T, T^\prime))
\end{equation}
For a given reaction temperature, inserting the respective $k$ back into equation~\ref{eq:product_form} yields the product concentration.
In addition, we can model the initial substrate in equation~\ref{eq:product_form} as a local parameter per batch~$S_{0, \mathrm{batch}}$.
Combined, this results in a hybrid model that yields product concentrations, which can be directly compared to the measured observations.
\\
In contrast, the reaction rate $k$ for a fixed temperature is first extracted by simple regression for the naive \ac{gp} approach, as it is common in biocatalysis.
It is then used to directly train a \ac{gp} to predict the influence of temperature.
The results of this approach and the hybrid model and a naive \ac{gp} are compared in Figure~\ref{fig:hybrid_model_pymc} (right panel), which shows the true function as well as samples from the two \acp{gp} to compare the mean and uncertainty predictions.
\\
The comparison shows how the hybrid model learns more efficiently from the seven observations provided per reaction, while the native \ac{gp} has a larger uncertainty.
In light blue, Thompson sampling is visualized as an acquisition function.
For the native \ac{gp}, the deviation of the mean from the true function is larger, which can best be seen around \SI{30}{\celsius}, where the observations of experiments from two different batches deviate.
The hybrid model can estimate the initial substrate concentration per batch and thus corrects for this bias.
Combined with the information from the reaction kinetics, this leads to an efficient model prediction.
Note that while replication of experiments is generally not recommended for \ac{bo} (Section~\ref{sec:practical_guidelines}), models quantifying batch effects or positional biases benefit from a few replicates of identical designs to quantify challenging experimental effects.
\\
In conclusion, this example highlights how using a hybrid model, consisting of a \ac{gp} in combination with mechanistic equations for reaction kinetics, can be more accurate and efficient for optimization problems with limited data availability and existing domain knowledge, which is often the case in biocatalysis and other domains of bioprocess engineering.

%% file: 09_Glossary/glossary.tex
\section*{Glossary}
\label{sec:glossary}
\paragraph{Acquisition Function} A mathematical function that guides the optimization process by quantifying the expected utility of evaluating a particular point, considering both the predicted value and uncertainty.

\paragraph{Batch} A group of experiments that are conducted in parallel or in sequence before the resulting observations are collectively analyzed and new experiments are proposed.

\paragraph{Design Variable} A controllable parameter or input variable in an experimental design that can be adjusted to optimize the outcome. In Bayesian optimization, these are the variables over which the optimization is performed and which form the parameter space (also referred to as search space or design space).

\paragraph{Experiment} The execution of an experimental procedure (wet lab or in silico) where specific design variable values are selected to obtain observations of system outputs. A single experiment may yield a single or multiple observations. For example, one may perform a one-time endpoint measurement of a single key performance indicator or record a time series of multiple values made at regular intervals.

\paragraph{Gaussian Process} A probabilistic model that defines a distribution over functions, specified by a mean function and a covariance (kernel) function, commonly used as a surrogate model in Bayesian optimization.

\paragraph{Heteroscedastic Noise} Heteroscedastic noise denotes that the variance of observation noise throughout the parameter space is not constant. A typical example in bioprocess engineering is the turbidity-based optical measurement of biomass concentration, where the observation noise increases for higher concentrations of biomass.

\paragraph{Hyperparameters} Specific parameters that control the behavior of an algorithm or model architecture, but are not learned from the data during training. In Bayesian optimization, these include structural elements like the choice of covariance function in a Gaussian process or pre-defined parameters like the length scale, which controls how strongly the surrogate model varies between observations. In contrast to model parameters, which are learned from training data, hyperparameters are specified separately before the training process.

\paragraph{Kernel} A kernel is a function that defines how similar or correlated two inputs are, typically based on their distance in input space. In \acp{gp}, the kernel governs how much information at one input point influences the prediction at another. Also referred to as a covariance function, the kernel is used to construct the covariance matrix that encodes prior assumptions about the smoothness, periodicity, or other structural properties of the function being modeled. This matrix underpins both the generation of sample functions from the \ac{gp} prior and the conditioning process used to make predictions after observing data. Hence, the kernel plays a central role in shaping the \ac{gp}'s behavior.

\paragraph{Machine Learning} A subdomain within Artificial Intelligence that focuses on algorithms that learn from (large) datasets to make predictions or decisions. In contrast to expert systems, which emulate human decision making by using fixed, pre-programmed rules of reasoning, Machine Learning focuses on improving model performance by directly learning from data. Moreover, Machine Learning emphasizes the mathematical and statistical foundations of learning algorithms.

\paragraph{Mixed-Integer Problem} An optimization problem where the parameter space consists of both continuous (e.\,g., temperature, concentrations) and discrete design variables (e.\,g., selection of organism or substrate).

\paragraph{Objective Function} A mathematical function that quantifies the performance or quality of a system configuration, which is to be optimized (maximized or minimized) during the optimization process. The objective function maps the design variables to a scalar value that represents the system performance, such as product yield, conversion efficiency, or process throughput.

\paragraph{Observation} The result of inferring a system output or key performance indicator based on one experimental measurement. Each observation is subject to noise arising from both inherent system variability and measurement uncertainties. An experiment can yield several observations.

\paragraph{Optimization Policy} A strategy that determines how to select the next point to evaluate in the optimization process, typically by balancing exploration of uncertain regions with exploitation of promising areas.

\paragraph{Parameter Space} The parameter space is the range of all possible input values (e.\,g., temperature, pH, concentration) in which the algorithm searches for the best result. It defines the limits within which the optimization process explores and evaluates different parameter combinations of the design variables.

\paragraph{Posterior Probability} Updated probability distribution after combining the observed experimental data with the prior beliefs according to Bayes’ theorem.

\paragraph{Prior Probability} Probability distribution in Bayes' theorem that describes the initial beliefs about a parameter before observing any data. In the context of Bayesian optimization, the distribution, also referred to as a prior, defines assumptions about the unknown objective function. These are subsequently updated with the experimental data, resulting in the posterior distribution.

\paragraph{Random Sampling} The process of generating random instances from a probability distribution, particularly important in Bayesian optimization for estimating uncertainty and exploring the parameter space.

\paragraph{Surrogate Model} A model approximating a system or function through the mapping of input data to its corresponding output. Surrogate models find application when evaluating the true system is costly, and the system is not fully understood. In Bayesian optimization, they approximate the true objective function and are iteratively updated with new data to guide the optimization process.

\paragraph{Uncertainty} Uncertainty represents the potential error or bias in a statistical estimate or measurement. Regarding (surrogate) models, uncertainty expresses the lack of knowledge in predictions, typically quantified through probability distributions or credible intervals in Bayesian optimization. Uncertainty in model parameters may arise from measurement noise or biases in the technical setup of an experiment, as well as from biological variability.